\documentclass[preprint]{aastex}

\shorttitle{Amplitude and mass-ratio distributions}
\shortauthors{Rucinski}

\begin{document}

\title{The photometric-amplitude and mass-ratio distributions 
of contact binary stars}

\author{Slavek M. Rucinski}
\affil{David Dunlap Observatory, University of Toronto\\
P.O.Box 360, Richmond Hill, Ontario, Canada L4c 4Y6}
\email{rucinski@astro.utoronto.ca}

\begin{abstract}

The distribution of the light-variation amplitudes, $A(a)$, in
addition to determining the number of undiscovered contact 
binary systems falling below photometric detection thresholds 
and thus lost to statistics, can serve as a tool in 
determination of the mass-ratio distribution, $Q(q)$, which is very
important for understanding of the evolution of contact binaries. 
Calculations of the expected $A(a)$ show that it
tends to converge to a mass-ratio dependent
constant value for $a \rightarrow 0$.
Strong dependence of $A(a)$ on $Q(q)$ can be used to determine
the latter distribution, but the technique is limited
by the presence of unresolved visual companions and by  
blending in crowded areas of the sky. The bright-star
sample to 7.5 magnitude is too small for an application 
of the technique while the 
the Baade's Window sample from the OGLE project may suffer
stronger blending; thus the present results are preliminary
and illustrative only. 
Estimates based on the Baade's Window data from the 
OGLE project, for amplitudes $a>0.3$ mag.\
where the statistics appear to be complete allowing
determination of $Q(q)$ over $0.12 \le q \le 1$,
suggest a steep increase of $Q(q)$ with $q \rightarrow 0$.
The mass-ratio distribution can be approximated by a
power law, either $Q_a(q) \propto (1-q)^{a_1}$ with
$a_1 = 6 \pm 2$ or $Q_b(q) \propto q^{b_1}$, 
with $b_1 = -2 \pm 0.5$, with a slight preference for the former 
form. While both forms would predict very large 
numbers of small mass-ratio systems,
these predictions must be modified by the theoretically
expected cut-off caused by a tidal 
instability at $q_{min} \simeq 0.07 - 0.1$.
A maximum in $Q(q)$, due to the interplay of a 
steep power law increase in $Q(q)$ for $q \rightarrow 0$ 
and of the cut-off at $q_{min}$, is expected to be mapped 
into a local maximum in $A(a)$ around $a \simeq 0.2 - 0.25$ mag. 
When better statistics of the amplitudes are available, 
the location of this maximum will shed light on the 
currently poorly known value of $q_{min}$.
The correction factor linking the apparent,
inclination-uncorrected frequency of W~UMa-type systems
to the true spatial frequency remains 
poorly constrained at about 1.5 to 2 times.
\end{abstract}


\section{INTRODUCTION}

Contact binary stars (also called W~UMa-type variable stars)
have a unique property among eclipsing binaries
in that their geometrical effects are much more important in defining
the shapes of their light curves than the atmospheric properties
of their components. Because of the similarity of effective
temperatures of the components, the eclipse depths are 
practically independent
of the effective temperatures, but depend strongly 
on geometrical parameters such as the orbital inclination, $i$, 
the degree of contact, $f$, and the mass ratio, $q$.  
This situation is very much different than for detached binaries
where -- in most cases -- differences in component
temperatures strongly affect the eclipse depths. Also, 
for a fixed mass ratio, 
just one parameter (the potential) defines the relative size 
of both components in a contact binary, in place of two 
independent radii as in detached binaries, so that
the light curves are described by fewer free parameters. 
 
The simplicity of the contact-binary light curves and lack 
of features in them (absence of external -- and
in most cases -- internal eclipse contacts) mean that
the light curves carry little information. 
Extensive calculations of large grids of light curves
\citep[Paper~I]{rci93} have shown that -- in the general case --
only two of the three main geometrical parameters are
well constrained by the light curve shapes. Only when radial-velocity
information on the mass ratio $q$ is available can the light curve
analysis also yield the $(i, f)$ pair and the whole set of the
parameters can be determined. From time to time one sees
attempts at determination of the mass ratio through analysis
of the $\chi^2(q)$ curve of the light-curve fit. This is
a very dangerous approach because such results on $q$ 
are very poorly constrained and, in fact, 
frequently plainly wrong, as the new radial
velocity data for many contact binaries 
\citep{ddo1,ddo2,ddo3,ddo4}
have shown: In numerous cases a spectroscopic 
(i.e.\ correct) value of the mass ratio is far away from 
the ``best'' photometric value of $q$, frequently far 
beyond the stated errors
estimated from the shape of the $\chi^2(q)$ curve. 
An exception to this indeterminacy is the special case
of the totally eclipsing contact systems. Such light
curves show characteristic inner contacts with duration 
of totality setting a strong constraint on the $(q, i)$ pair
\citep{mki72a,mki72b}.

Instead of considering results of individual light curve solutions, 
augmented by radial velocity studies, one can attempt to utilize
statistical distributions of the observed parameters
for large samples of binaries. The simplest
to obtain and most obvious among such distribution is the 
photometric amplitude distribution. However, it has never been 
used, probably because it is generally recognized that 
such an observational distribution must be very closely 
related to (if not observationally defined by) discovery 
selection effects. Now the situation is changing:  
New data from microlensing surveys and from massive, deep
sky surveys, aimed at other goals, but extensive and 
statistically rigorous, start providing a wealth of sound 
``by-product'' information for contact binaries
\citep{ogle1,ogle2,macho,ogle3,ogle-cl,LMC,ogle4,ogle5}.
Such surveys, with well controlled selection biases, 
are far superior to any statistical inferences based on
individual light and radial velocity solutions as the
latter are very heavily biased by discovery and observational
selection effects and by observer preferences.

This paper is basically a much expanded version of a preliminary
discussion in Section~5 of \citet{ogle2}. It attempts to
characterize and analyze information contained in
light variation amplitudes of contact binaries.
By definition, we call the amplitude the depth -- in magnitudes --
of the deeper eclipse.
The amplitude distribution is considered here as a tool to study
two important issues: (1)~How many contact systems are missed
and remain unknown due to their
amplitudes being smaller than the detection threshold,
and (2)~Can such a distribution be used to derive properties of contact
systems, especially the distribution of the mass ratios? Both
questions can be addressed for the idealized case of isolated
contact binaries. However, because of the blending problems
which affect the currently-best statistics for the Baade's
Window data and -- even more importantly -- 
because contact binaries frequently
have close, unresolved companions, the results of this
paper must be considered as preliminary and illustrative. 

The paper continues the approach of Paper~I
in that we analyze the main, global properties of the 
light curves by covering the entire
range of relevant parameters. While the approach may appear 
to be coarse, the intention is to uncover the main relations 
and dependencies. We feel that this is justified in view 
of lack of similar global approaches in this field. 

We give an expanded background in 
Section~\ref{background} while the results of 
the calculations are presented in Section~\ref{results}.
Applications of the theoretical results to observational statistics
based on the data for the OGLE Baade's Window sample, to derive the
mass-ratio distribution,
are discussed in Section~\ref{mass-ratio}. This distribution
cannot yet be analyzed at its low mass-ratio end because
of the poor statistics for small-amplitude systems. However,
we predict that the expected cut-off to the 
mass-ratio distribution at $q_{min}$
will strongly modify the amplitude distribution at its low end; 
the matter of the smallest possible mass ratios, $q_{min}$, 
is discussed in Section~\ref{min_q}. The results of the
paper and their implications for our understanding of the
evolution of contact binary systems are discussed in Section~\ref{evol}.
The conclusions of the paper are summarized in Section~\ref{concl}.

\section{BACKGROUND ON PREDICTION OF AMPLITUDE DISTRIBUTIONS}
\label{background}

\subsection{The contact binary model}

Current light-curve synthesis models permit computation of light
curves with almost arbitrarily high accuracy. The accuracy
depends basically only 
on how many integration points one is ready to use. Obviously,
the model must represent real contact binaries, a requirement 
which -- with the improved numerical accuracy of the models
-- is becoming increasingly difficult to fulfill
in view of many complications such as 
activity-induced star spots or possible deviations 
from plane-parallel atmosphere models in the ``neck'' 
region between components of a contact binary. 
In this paper, we neglect all complications
of spots and stellar activity and concentrate on 
the basic dependencies of light-variations
on the geometrical parameters. In order to do so,
we fix the atmospheric parameters, such as the limb and gravity 
darkening coefficients. It may be argued that they are relatively
less important than the geometrical parameters because 
-- due to constancy
of the effective temperature over the whole common surface -- 
these coefficients are practically the same everywhere. 
Thus, as in Paper~I,
we consider only one representative wavelength (matched to the
$V$-band filter) and one effective temperature for
a typical solar-type star (5770~K). Extensive tests have shown that
such a combination gives reasonable representation of the observed
light curves of contact binaries for yellow-red spectral 
region (in this
paper we utilize observational amplitude distributions from the OGLE
sample in the photometric $I$ band) and for a wide range of
effective temperatures at which the contact binaries are
most common, i.e.\ with spectral types from early F through G to early K.

The geometrical elements which define the shapes of the contact-binary 
light curves are designated as $(i, f, q)$ \citep{book1,book2}. 
The orbital inclination, $0 \le i \le 90^\circ$, 
is conventionally defined
in such a way that $i=0^\circ$ corresponds to the orbit in the
plane of the sky and
$i = 90^\circ$ corresponds to the orbit seen edge on.
The simplest assumption concerning the orbits is that they
are randomly oriented in space. 
This means that probability of detection of a binary with a 
given $i$ is proportional to $\sin i$. This is because for 
$i \rightarrow 0^\circ$, the available 
range of orientations of the orbital ascending node goes to zero.
The degree of contact is usually defined through the
potential of the common equipotential surface. In this paper we 
consider only three discrete values of the degree-of-contact
parameter, $f = (C_1 - C)/(C_1 - C_2)$, where $C$ is the 
Jacobi constant for a common equipotential: 
$f=0$, $f=0.25$ and $f=0.5$. The inner contact
with the components just touching corresponds to $f=0$. The
frequency distribution of the parameter $f$, $F(f)$, is currently 
unknown. However, analysis of a large number of contact
systems in the OGLE sample \citep{ogle2} has 
confirmed the earlier indications \citep{luc73,rci73}
that $f$ is typically small, within $0 < f < 0.5$, 
with a tendency for a (poorly defined) maximum around $f = 0.25$. 
Large values of $f \rightarrow 1$, corresponding to the outer
contact, are not observed. 

The third geometrical parameter is the mass ratio. From the
point of view of the contact binary formation and evolution this is 
astrophysically the most important parameter among those
that define the light curve shape. We define the mass ratio 
as $q = M_2/M_1 \le 1$. From many radial velocity studies -- which 
tend to preferentially favor equal-mass combinations --
we know that contact systems definitely avoid
the equal-mass situation of $q=1$ and this fact has a good
theoretical explanation (see Section~\ref{evol}).
Systems with $q$ approaching unity do exist; an extreme 
example is the recently discovered system
with very good radial velocity data, V753~Mon, which
has $q=0.970 \pm 0.003$, \citet{ddo3}. However
-- in spite of the relative ease of detection of large
mass-ratio systems -- we see very few systems with $q > 0.8$,
so  they must be very rare in space. 
Currently, the only property  of the mass-ratio distribution, $Q(q)$,
that we can be sure of is its 
tendency to increase for $q \rightarrow 0$, but the rate
of the increase is not known. This paper attempts to determine
$Q(q)$ in Sections~\ref{results} -- \ref{min_q}; 
the results will be related to the theoretical evolutionary
predictions in Section~\ref{evol}. 

As the reader may have already noticed, we use capital letters to denote 
the distributions of parameters denoted by lower case letters.
In particular, the amplitude distribution
can be written as $A(a)\, da = (dN/da) \, da$ so that $A(a)$ takes the
significance of the number of binary systems observed in the elemental
increment $da$. 

\subsection{Derivation of the amplitude distribution from light
curve synthesis models}

The amplitudes predicted from a light-curve 
model can be written as three-dimensional functions: 
$a = a(i, q, f)$. In practice, we use arrays calculated
for fine grids of the parameters $i$, $q$, $f$. 
To simplify handling of such large arrays, 
and because we know that the degree-of-contact is not very
important in defining the light curve shapes, we have
computed such arrays for only a few fixed values of $f$. 
Each amplitude distribution for a fixed value
of $q$, $A_q(a)$, can be also considered individually; such
normalized distributions will be used in an 
attempt to derive the mass-ratio distribution $Q(q)$
(Section~\ref{mass-ratio}). For the cases
in which  discrete values of $f$ and $q$ are used, we will use the
notation $a_q^f$. In particular, we will consider the
functional dependence of the amplitude on the orbital
inclination, $a_q^f = a_q^f (i)$ as well as the distribution of the
amplitudes, $A_q^f = A_q^f(a)$. 

We derive the expected amplitude distribution, 
$A_q^f(a)$, by eliminating the orbital-inclination
dependence under the assumption of random orientation of the 
orbital planes in space. Two approaches are possible, both 
utilizing light curves computed with the light curves synthesis codes.
One is a semi-analytical approach:
We start with the dependence $a = a(i)$ computed by 
varying the inclination in the models (for a fixed combination of
$q$ and $f$). For a random distribution
of orientations, the number of discovered systems should scale 
with the inclination angle $i$ as:
$dN = \sin i \, di$, 
which can be written as
$(dN/da)\,(da/di) = \sin i$, hence
\begin{equation}
A(a) = (dN/da) = \sin i(a)\, (da/di)^{-1}   \label{eq1}
\end{equation}
The amplitudes monotonically increase with $i$ so that there is no
difficulty in inverting the function $a(i)$ into $i(a)$. However,
the reciprocal of the derivative $da/di$ must be computed
numerically. This is a complication because the light-curve synthesis
results on $a(i)$ carry numerical noise which becomes amplified in
the numerical differentiation. 

\placefigure{fig1}

We can get some insight into the behavior of $da/di$ by
analyzing one specific case, as shown in Figure~\ref{fig1}.
The left panel of the figure 
shows 45 light curves computed with inclinations
varied in 2-degree steps for a representative case of 
$q=0.35$ and $f=0.25$. The corresponding variations of $a(i)$
and $da/di$ are shown in the right panel. 
Notice the progressive increase of the amplitude
with inclination to about $i \simeq 70^\circ$ where the derivative
has its peak and then flattening of the $a(i)$ curve
in the region of the total eclipses. From the shape of the
$(da/di)$ curve, we can expect
two regions at both ends of the distribution 
where $(da/di)^{-1}$ will be particularly 
difficult to evaluate, one for $i \rightarrow 0^\circ$ and one for
$i \rightarrow 90^\circ$. The region at $i \rightarrow 90^\circ$
is usually small, but its size depends on the mass ratio and
grows for small values of $q$. This region is populated by systems 
showing total eclipses of similar depths for orbital inclinations
close to $90^\circ$. At the other end, for $i \rightarrow 0^\circ$, 
the light curves have very small amplitudes which change very little
with $i$. The resulting increase $(da/di)^{-1} \rightarrow \infty$ 
is expected to be moderated in Equation~\ref{eq1} by the term 
$\sin i \rightarrow 0$, so that the result is difficult to predict. 
We show later on in Section~\ref{results}
that -- somewhat unexpectedly -- for $a \rightarrow 0$, 
the amplitude distributions for all
values of $q$ appear to converge to mass-ratio dependent 
constant values.

A much simpler second approach to evaluate $A(a)$,
which avoids the numerical difficulties of computation of
$(da/di)^{-1}$,
is through a Monte Carlo simulation. In such an experiment,
the inclinations are drawn randomly with a 
distribution $I(i) \propto \sin i$. These are then
interpolated in the amplitude arrays $a_q^f = a_q^f (i)$, to form the
amplitude distributions by simply binning the resulting distributions,
$A(a) = (dN/da)$. This was the approach actually used in this paper.
It has the advantage of direct modeling of 
the difficult regions at $i \rightarrow 0^\circ$ and 
$i \rightarrow 90^\circ$, and entirely avoids 
evaluation of $(da/di)^{-1}$.

\placefigure{fig2}

We show in Figure~\ref{fig2}
a representative case of the amplitude distribution $A(a)$ 
computed using the Monte-Carlo approach for the same light curves as in
Figure~\ref{fig1}, that is for the mass ratio $q=0.35$ and 
for the degree-of contact parameter $f=0.25$ (this figure shows
also the corresponding distributions for $f=0$ and $f=0.5$).
We will discuss an amplitude distribution like 
this one in the next Section~\ref{results},
after summarizing the details and assumptions of the Monte Carlo
computations.

\section{CALCULATIONS OF THE AMPLITUDE DISTRIBUTION}
\label{results}

\subsection{Details of the Monte Carlo calculations}

For consistency with previous calculations of the large grid of
light curves in Paper~I, the light-curve grid used to 
predict the amplitude distributions utilized the same assumptions on
the temperatures, limb and gravity darkening laws and relative fluxes.
We suggest the reader consult that paper for details. The only
differences are as follows: (1)~We improved the sampling of the 
degree-of-contact parameter, $f$, by adding 
the most likely value of $f = 0.25$, in addition 
to the two bracketing values considered before, $f = 0$ (the inner
contact) and $f=0.5$; we have also dropped the unobserved case of
$f=1$ (the outer contact); 
(2)~The amplitude calculations have been extended
over the whole width of the inclination angles, 
$0^\circ \le i \le 90^\circ$
(the previous calculations applied only to $i \ge 30^\circ$);
(3)~While the parameter $f$ has been sampled rather
coarsely, the parameters $q$ and $i$ have been sampled
with fine grids of $\Delta q = 0.01$ and $\Delta i = 1^\circ$;
(4)~The amplitude distributions have been computed using 
the magnitude scale (note that the light intensity was used 
in Paper~I), with a fine grid of $\Delta a = 0.01$. Later on,
for practical applications, larger bins of $\Delta a = 0.05$ 
have been used.

In the Monte Carlo experiments, the inclinations were drawn 
using uniformly distributed random numbers in the 
$x \in [0,1]$ range, mapped into the $\sin i$-distributed 
inclination angles through $i = \arccos x$. Each experiment
for a fixed pair of $(f,q)$ involved $4 \times 10^6$ samples of 
$a_q^f = a_q^f(i)$ which were then binned into $A_q^f(a)$ distributions
over $0 \le a \le 1.1$, with 110 bins of $\Delta a = 0.01$. 
Thus, typically, each bin $\Delta a$ contained $4 \times 10^4$ samples
giving an error at the level of about 0.5 percent.
Results of more accurate computations of the normalized $A_q(a)$ with 
larger bins of $\Delta a = 0.05$ and for 10 values of the mass ratio 
in intervals of $\Delta q = 0.1$
are given in Table~\ref{tab1} for the case of $f=0.25$. 
The wider bins in $a$ and $q$ result in an increase of 
accuracy by a factor
of about 4 times so that the $A_q(a)$ distributions 
in Table~\ref{tab1} should be accurate to about 0.1 percent. 
The distributions for $f=0$ and $f=0.5$, have been also computed,
but we do not give them here for economy of space\footnote{The 
tables similar to Table~\ref{tab1} for $f=0$ and $f=0.5$, as well
as the detailed tables sampled at $\Delta q=0.01$ and $\Delta a =0.01$ 
are available over the Internet from 
http://www.astro.utoronto.ca/$\sim$rucinski.}.

\placetable{tab1}

Figure~\ref{fig2}
shows the detailed behavior of the amplitude distributions 
$A_{0.35}^{0.0}(a)$, $A_{0.35}^{0.25}(a)$, $A_{0.35}^{0.5}(a)$, 
i.e.\ for a case of $q=0.35$, and for $f=0$, 0.25, 0.5.
The narrow interval of $\Delta q = 0.01$ around $q=0.35$
has been used in the figure for illustrative purposes because this way
we can clearly demonstrate the narrow peak of $A(a)$
at large amplitudes which corresponds to the range of inclinations giving 
total eclipses. When larger bins in $q$ are considered,
as in Figure~\ref{fig3} (see the panel labeled ``0.35'' for 
a direct comparison with Figure~\ref{fig2}),  
the peak becomes broader and less prominent.
Later on, in actual applications in Sections~\ref{mass-ratio}
and \ref{min_q}, we use $\Delta q = 0.1$ or $\Delta q = 0.2$.
The location of the total-eclipse 
peak weakly depends on the degree-of-contact,
$f$: The amplitudes are larger for contact systems with stronger
contact and smallest for binaries with components just touching.

\subsection{Dependence of the amplitude distribution on the mass ratio}

Figure~\ref{fig3} shows the ten predicted amplitude 
distributions $A_q^{0.25}(a)$, sampled at $\Delta a = 0.05$, 
for $f=0.25$, with $q$ spanning the whole
range $0 < q < 1$ in intervals of $\Delta q = 0.1$.
The individual panels of the figure 
correspond to columns in Table~\ref{tab1};
each $A_q(a)$ has been separately normalized to unity.
Note the same details as for the specific case 
of $A_{0.35}(a)$ in Figure~\ref{fig2},
but now displayed for the whole range of the mass ratio, in wider
bins in $q$ and $a$. The figure illustrates the strong dependence
of $A_q(a)$ on the mass ratio. While all fixed-$q$ distributions
show a rise towards small amplitudes, the range of the
amplitudes is very different: For large $q$, all amplitudes are
possible, while for small $q$, only small ones are permitted. 
If one particular value of the mass ratio were to dominate in the
mass-ratio distribution $Q(q)$, the characteristic features of
its corresponding $A_q(a)$ would be present
in the combined $A(a)$.  
The observed amplitude distributions appear to be featureless,
so that the corresponding mass-ratio distributions must be
smooth, too. As we discuss in Section~\ref{mass-ratio}, 
the only characteristic feature
that we can be sure of is lack of large amplitude systems 
indicating a strong weighting towards small mass ratios. 

\placefigure{fig3}

In addition to all individual distributions of $A_q(a)$ in
Figure~\ref{fig3}, 
we also show distributions resulting from a simple addition
of all distributions over the whole range of $q$. 
This, in effect, corresponds to the
case of $Q(q) = const$. Such a distribution also monotonously
increases towards low amplitudes, as do the individual
$A_q(a)$ distributions. The local peaks at the
characteristic maximum amplitudes in individual $A_q(a)$
disappear in the combined $A(a)$ through ``dilution''
when summed with other distributions. Bearing in
mind the independent normalizations of 
each $A_q(a)$ and for the combined distribution, 
we can see that all $A(a)$ distributions
look very similar in the central portions of the 
amplitude range and show the strongest dependence of the shape 
on $q$ at low and large values of the amplitudes.

\subsection{The detection threshold}

Of particular importance to practical applications of our calculations
is the behavior of $A_q(a)$ for small amplitudes,
because this determines how many systems would be lost in 
photometric surveys 
with specific detection thresholds. As we can see in 
Figure~\ref{fig2} for the case of $q=0.35$,
the amplitude distribution tends to converge to constant value,
$A(a \rightarrow 0) \simeq 0.02 - 0.03$, per the $\Delta a = 0.01$
bin. The same convergence is observed for the wider bins used
in Figure~\ref{fig3}, for all values of $q$. Thus, it appears that,
for $i \rightarrow 0^\circ$, the $\sin i$ term 
in Equation~\ref{eq1} does not force
$A(a) \rightarrow 0$ for $a \rightarrow 0$. In practical terms,
it means that the first bin in any amplitude histogram, 
$[0, \Delta a]$, 
will always contain some objects, irrespectively of the
size of the bin. This is an important
new result indicating that a substantial number of contact binaries
may have small amplitudes below detection thresholds. For the
specific case of $q=0.35$ illustrated in Figure~\ref{fig2}, if the
threshold were at the (currently unachievable)
$a_{min} = 0.01$ mag., then the loss would be about
2 -- 3 percent.

\placefigure{fig4}

Figure~\ref{fig4} shows -- as a function of $q$ -- the fraction 
of systems falling below an assumed value of the threshold amplitude,
$\Sigma A_q(a < a_{min})$. For the 
detection threshold of $a_{min}=0.01$ mag., only about 2 to 5 percent of
systems would be typically lost; this can be verified by inspecting 
the first bin $0 < a < 0.01$ in Figure~\ref{fig2}. 
However, Figure~\ref{fig4} indicates that
for a more representative value of
0.1 mag., some 20 to 40 percent would remain undetected; 
for larger $a_{min}$ the fractional loss would be larger. 
Note that for small values of $a_{min}$, 
there exists a simple proportionality between the
fraction of the undetected systems and the height of the
threshold, $\Sigma A_q(a < a_{min})
\propto a_{min}$, which is a direct
consequence of the convergence of $A_q(a)$ to a constant value
for $a \rightarrow 0$.

The detection losses strongly depend on the mass ratio. Thus,
the main question is, what are the mass ratios and how they 
distribute the losses due to the 
detection thresholds? As we will see later, the
mass-ratio distribution appears to be 
steeply rising for $q \rightarrow 0$ which makes
the left hand edge of Figure~\ref{fig4} particularly important
for estimation of the fraction of undetected systems.

\subsection{Maximum amplitudes}

As one can see in Figure~\ref{fig3}, for each mass ratio, 
there exists a characteristic maximum amplitude, $a_{max}$.
This maximum amplitude grows with $q$.
The mere presence of large amplitudes in an observed distribution
indicates that among contact binaries of the sample 
there exist ones with large mass ratios. 
 
The maximum amplitudes are tabulated as a function of $q$ 
for the three values of the fill-out parameter $f$ 
in Table~\ref{tab2}. They are also shown
in Figure~\ref{fig5}. We will discuss in the next section that
the well established and most trustworthy part of the observed $A(a)$ 
extends only above $a \simeq 0.3$. As a result, nothing
can be said about the mass-ratio distribution below $q \simeq 0.12$.
This inter-relation between $a_{max}$ and the accessible range of 
mass ratios emphasizes the importance of the good detection 
statistics at low amplitudes. We will see later in
Section~\ref{min_q} that the definition of $A(a)$ in the region
around $0.1 < a < 0.3$ is particularly important for shedding light
on the expected low mass-ratio cut-off in $Q(q)$ at 
$q_{min} \simeq 0.07 - 0.1$.

\placefigure{fig5}

\placetable{tab2}

\section{DETERMINATION OF THE MASS-RATIO DISTRIBUTION}
\label{mass-ratio}

\subsection{Observational data: 7.5 magnitude-limit sample}

The strong dependence of the amplitude distributions 
on the mass ratio can be utilized to derive the
functional shape of $Q(q)$ for the observed systems. The most
obvious source for $A(a)$ are the catalogue data for contact
binaries, as listed in the General Catalogue of Variable Stars 
\citep[GCVS]{gcvs}.
However, this material must be used judiciously as
it is known to be strongly affected by discovery and observational
selection biases. We used the most recent, October 2000 
version of the GCVS, which is only available 
electronically\footnote{The GCVS has been obtained from
ftp://ftp.sai.msu.su/pub/groups/cluster/gcvs/iii/.}. It
consists of the main catalogue of three volumes, augmented
by the ``name lists'' number 67 to 74. We added 
name list number 75 from the same source and cross-referenced
the combined catalogue with the list of variable stars in the
Hipparcos catalogue \citep{hip}. At that point, the variability
types were checked on the basis of the Hipparcos light curves
and several incorrect or disputable assignments of types were found.

Our previous experience with the amplitude statistics of contact
binaries, based on the GCVS data \citep{RK94,ogle2},
indicated a strong dominance of large amplitude variables
for the simple reason that large amplitude 
variables are always easier to discover than small amplitude ones. 
Such an excess of large amplitude systems seemed 
implausible, even without confrontation with the
results of this paper which predicts dominance of small
amplitudes. To avoid the bias of only large-amplitude
systems being discovered among fainter systems, 
an assumption has been made that the sky has been fully
inspected for variability by numerous observers and by
Hipparcos to a relatively bright level. This level
has been selected to coincide with the astrometric
completeness limit of the Hipparcos mission at
$V \simeq 7.5$. The limiting minimum amplitude for such a sample
is unknown.  
Perhaps it is $a_{min} \simeq 0.1$, although fainter systems were
detected down to 0.03 mag. We feel
that this is the most complete currently-available 
sample for the sky-field contact binaries. Deeper samples
that we attempted to construct in the same way indicated
the presence of discovery selection losses past $V \simeq 7.5$. 

An important limitation in the current context is the presence
of unresolved physical companions of contact binaries leading
to photometric blending and a decrease in observed amplitudes. 
This is a difficult and murky area as no statistics exist
to evaluate importance of the blending. Some observers remarked
that close companions surprisingly frequently accompany
W~UMa binaries \citep{RK88,cha92,HM98}. We also keep discovering 
them in our spectroscopic survey of short-period binary systems
which is currently conducted at the David Dunlap Observatory
\citep{ddo1,ddo2,ddo3,ddo4}. The spectroscopic detection 
is the most sensitive and least biased of all 
available techniques, although speckle interferometry
surveys have been already 
more systematic in surveying all bright stars
of the sky. The spectroscopic analysis is particularly
easy when use is made of broadening-functions
\citep{ruc99} which permit separation of 
components even for rather large difference in brightness
through very different spectral signatures of broad and
sharp components \citep{ddo4}. 

A ``third light'' contribution relative to the maximum brightness
of the contact system, $x = L_3/(L_1+L_2)$, is expected to change the
true amplitude $a$ to the observed one
$a' =-2.5 \log[(10^{-0.4\,a}+x)/(1+x)]$. 
At present it is very difficult to quantify
the influence of the unresolved companions 
on the observational amplitude distribution 
because we have no idea about frequency
of occurrence of the companions and thus about the distribution
of the quantity $x$. If contact binaries are formed 
preferentially in a hierarchical process which produces
wide orbits first and leaves pre-stellar
clouds of low angular momentum to form contact systems,
the frequency of occurrence may be higher than for other stars.
Further, if the close binary formation process has more to do with
random pairing then $x$ should be -- on the average -- small, but
if the process tends to prefer equal-mass components, then 
$x \simeq 1$. We see contact binaries in systems with bright
companions, such as 44~Boo, with companions comparable
in brightness as in HT~Vir, but we also see faint, low
mass companions to systems like VW~Cep, so that apparently
all values of $x$ are possible. The nearby systems
offer the best chance of detection of the companions so
that the Hipparcos sample is probably the best one to start
from.

We have created what we call the ``7.5 magnitude limit'' sample
of contact binary systems
on the basis of the merged GCVS and Hipparcos data for binary
systems with periods shorter than one day. By utilizing
partially unpublished spectroscopic results from the
David Dunlap Observatory, we
could clean the sample of short-period pulsating stars and
take into account the presence of close companions. 
The full discussion of the sample, which consists of 41 close 
binary systems, in that 10 systems having visual, speckle or
spectroscopic companions, will be 
a subject of a separate investigation. 
We summarize here only the preliminary conclusions based on this
sample concerning the amplitude distribution.

The sample includes all binaries
designated in catalogues by codes ``EW'', ``EB'' or ``Ell'' to the
limiting maximum magnitude of 7.5 mag.\
and with periods shorter than one day. While we wanted to
isolate true contact binaries, their separation from the
related semi-detached and poor-thermal-contact
systems was not easy. For that
reason we initailly considered the EW and EB groups together,
recognizing that the semi-detached EB systems 
are on the average brighter than contact systems
so that we see them deeper in space; therefore,
the EB systems are over-represented 
in a magnitude-limited sample such as the one considered here. 
As a first step, we carefully checked
the light-curve types and in a few cases exchanged the EW and EB
types. Since most
systems do not have radial velocity data and the depth-of-eclipse
criterion does not always give a unique answer, the separation
of systems into the two groups 
remains preliminary. The ``Ell'' (ellipsoidal)
variables were also included because they are
mostly a mixture of the EW or EB systems just seen at
low orbital inclination angles. We made an effort to assign
them to either EW or EB groups on the basis of 
light-curve shapes and relative
depths of eclipses -- equal or unequal -- 
respectively. The final numbers are 27 EW systems and 14
EB systems. Only 13 EW and 10 EB systems have retained the
classification as in the Hipparcos catalogue.
An additional complication in a separation of the two groups of 
binaries is the fact that some apparently genuinely
contact systems show unequally deep eclipses. There are very
few such poor-thermal-contact
systems, about 2 percent in the volume-limited OGLE
sample \citep{ogle2}, but they do have deeper primary eclipses
than well behaving contact binaries and would normally be
classified as the EB systems.

\placefigure{fig6}

The EB systems appear to have -- on the average -- 
longer periods than EW systems so that
different period distributions for both groups can help 
in assigning the type and separating the two groups. In particular, 
contact binaries are extremely rare in volume-limited
samples for periods longer than about 0.6 -- 0.65 day
(see Figure~1 in \citet{ogle-cl}).
The left panel of Figure~\ref{fig6} 
shows the amplitudes plotted versus the orbital period
for the whole sample of 41 systems with $P < 1$ day,
with different symbols for EW and EB groups.
With the additional constraint of $P < 0.65$ days,
the 7.5 magnitude-limit sample shrinks by about one half, but
presumably consists mostly of contact binaries; it consists of
20 EW systems and 4 EB systems (among the latter, two with
very small and two with moderate amplitudes, hence not
really typical for the EB's). 
Figure~\ref{fig6} shows how (still uncertain)
light contributions from close companions affect the
observed amplitude distribution. 
While the companions are well known in frequently observed
systems such as 44~Boo, HT~Vir or VW~Cep, 
not all stars of the 7.5 magnitude-limit sample
have been scrutinized for presence of
companions so that the amplitude distribution for the
subsample with $P < 0.65$ days, shown in
the right panel of Figure~\ref{fig6}, must be considered 
as preliminary.

The 24 systems with $P < 0.65$ days
form too small a sample to securely define the
amplitude distribution for the mass-ratio determination.
Since it is quite unlikely that the 7.5 magnitude-limit sample
is missing bright contact systems with amplitudes larger than
$a \simeq 0.1$, an increase in numbers for better 
statistics could be now achieved by deeper systematic 
searches for contact sky-field systems.  
(Note, that according to the previous
Section, the total loss of ``zero-amplitude'' or $a < 0.1$ 
systems is still substantial at about 20 to 40 percent.) 
Currently, 
an extension beyond the 7.5 magnitude limit on the basis of
catalogue data would be too risky to attempt: From
among the 41 systems of the sample, as many as 17 (i.e. 41
percent) have been
discovered by the Hipparcos mission which is complete only
to $V \simeq 7.2 - 7.8$. There exists no other deeper 
all-sky survey which would compete with the Hipparcos survey in terms
of the systematic temporal coverage and photometric accuracy.

Despite the small size of the 7.5 magnitude-limit sample,
we can note in Figure~\ref{fig6} the absence of
large-amplitude systems (except for
HT~Vir, but only after its amplitude is corrected
for the presence of its companion) and the well-defined
rise of the amplitude distribution toward the
small amplitudes, as expected by our results in
Section~\ref{results}. 

\subsection{Observational data: The OGLE sample}

The results of the OGLE project 
\citep{ogle1,ogle2}, more fully interpreted in \citet{ogle-cl},
provide sound data on the 
observed amplitude distribution, $A_{obs}(a)$. 
The statistics are based on two volume-limited 
samples, to $d = 3$ kpc and to $d = 5$ kpc, designated
as BW$_3$ and BW$_5$. As discussed in \citet{ogle-cl},
the samples are complete to the absolute magnitudes $M_V = 5.5$ and
$M_V = 4.5$, respectively. The sample sizes  
are 98 systems for BW$_3$ and 238 systems for BW$_5$, with
BW$_5$ including BW$_3$. There 
may exist a dependence of the amplitude on the 
absolute magnitude: The hotter, brighter systems may show a stronger
admixture of EB binaries. Since BW$_5$ consists preferentially
of brighter systems seen deeper in space, while BW$_3$
is based on fainter, more local systems, it was felt prudent
to consider the two samples separately.
The BW$_3$ sample would be in general
the preferred one as it is expected to better represent
the typical population of contact systems. 

The main limitation of the statistics based on the OGLE data
is blending of the images in the crowded Baade's Window
area, leading to systematically smaller 
variability amplitudes. The random-pairing blending occurs 
on top of the influence of close companions, as for the 
nearby stars. The difficulty is that in the case of the
OGLE survey the stars are
not easily accessible to medium-resolution spectroscopy which
would not only provide confirmation of binarity (i.e.\ elimination
of $\delta$~Sct and RR~Lyr stars), but would also
permit detection of spectroscopic companions. Since we
cannot address the matter of blending, we ignore it entirely.
We suggest that our analysis of the OGLE data should be
taken as an illustration how blending-corrected data could  
normally be treated using our approach.

\placetable{tab3}

\placefigure{fig7}

The observed amplitude distributions $A(a)$ 
for BW$_3$ and BW$_5$ are tabulated in Table~\ref{tab3}
and are shown in Figure~\ref{fig7} by shaded
histograms. The distributions are given with
the amplitude bins of $\Delta a = 0.05$, centered on the values given
in the first column of Table~\ref{tab3}. 
The amplitudes are in the photometric $I$-band. Because the
$I$-band amplitudes are typically only 3--5 percent 
shallower than in the
$V$-band, we disregarded the small difference which is immaterial
in view of the current, poor statistics. However, the matter of the
band matching may have to be addressed in future, by more advanced
investigations.

The completeness threshold for discovery 
of the OGLE sample was estimated in \citet{ogle1} 
at about 0.3 mag. As stated in Section~\ref{results},
the basis for this estimate was the
absence of a further increase in numbers
of detected systems for $a < 0.3$ mag. The assumption that
the OGLE sample is complete for $a > 0.3$ may be
conservative, but provides full assurance of an unbiased 
statistics of the amplitudes. Also, even when measurement errors 
($\sigma$) 
are at the level of 0.01 -- 0.03 mag., as for the OGLE project,
detection of variability requires a signal several times larger,
say $5\,\sigma$. To characterize the variability and estimate
the variability type requires still more margin. All in all,
the full completeness limit of 0.3 mag.\ is not at all unrealistic
in such a case. The OGLE project has in fact discovered several
contact binaries with $0.1 < a < 0.3$, but we suspect that not
all contact systems have been discovered in this interval. 

The continuous lines in Figure~\ref{fig7} give the predicted
amplitude distribution $A(a)$ calculated assuming $f=0.25$ and
a flat distribution $Q(q) = const$ (this is the same
as the one marked by the thin line in Figure~\ref{fig4}),
superimposed on the OGLE distributions.
Absence of large-amplitude systems in $A_{obs}(a)$ is striking. 
This may be because the mass-ratio values 
close to unity are extremely rare ($Q(q)$ rising for
$q \rightarrow 0$) or because of strong blending of images or
-- most likely -- both. Since we cannot estimate the blending 
effects, we assume that the shape of $Q(q)$ is reflected in 
$A(a)$. The results of this assumption are described below.

\subsection{Determination of the mass-ratio distribution}

An observed amplitude distribution can be modeled by adjusting
the $Q(q)$ distribution. We can predict the $A(a)$
distribution by utilizing the computed distributions 
$A_q(a)$ (Section~\ref{results}):
\begin{equation}
A_{pred}(a) = \Sigma Q(q_i) A_{q_i}(a)   \label{eq2}
\end{equation}
Strictly speaking, we should denote the fact 
that we use a specific value of $f$,
so that a proper superscript would be in order. For clarity,
in what follows, we will assume $f=0.25$ unless noted otherwise.
The functions $A_{q_i}(a)$ are each normalized to unity (for
each interval of $q$); this
permits expressing $A_{pred}(a)$ and $Q(q)$ in the actual numbers
of systems so that uncertainties can be simply estimated from
the Poisson statistics. 
We attempted to determine $Q(q)$ by representing it by
five independent bins $\Delta q = 0.2$ wide and by 
two-parameter power-laws, as described below.

\subsubsection{Random-search fits}

As the first step in estimating $Q(q)$ on the basis of the
OGLE amplitude distribution,  
a simple random search for a best fit of $A_{pred}$
to $A_{obs}$ was conducted by using five bins
of $\Delta q = 0.2$. Each bin of $Q(q_i)$ was considered independent,
without any assumption of smoothness or continuity constraints
on $Q(q)$. The solution was obtained by an extensive random trial
search, iterated until the smallest value of 
the ``quality-of-fit'' measure, $\chi^2$, 
defined as $\chi^2 = \sum (A_{obs} - A_{pred})^2/\sigma_A^2$ was found.
Poissonian estimates $\sigma_A = \sqrt{A_{obs}}$ 
for each bin were used for the standard errors.
Because of the small number of filled bins in $A_{obs}$ for $a>0.3$
(7 and 8 for BW$_3$ and BW$_5$), the 5-parameter description of $Q(q)$
obviously could be considered only as indicative, yet perhaps useful
as the first stage.

\placefigure{fig8}

The detailed results on $Q(q)$, expressed as the number of systems
per a bin of $\Delta q = 0.2$, are given numerically 
and graphically (continuous line histogram) in Figure~\ref{fig8}. 
Note that the first
bin $0 \le q \le 0.2$ senses the amplitude distribution only
between our low limit of $a=0.3$ and $a \simeq 0.43$ which
corresponds to $q=0.2$. Nevertheless, this is the region where
the observational $A(a)$ for the OGLE samples were best determined.

While the main solution was done with $A^{0.25}_q$, 
i.e.\ for the case of $f=0.25$, as tabulated in Table~\ref{tab1}, 
we made also fits for $f=0$ and $f=0.5$ ($A^{0.0}_q$ and $A^{0.5}_q$).
These solutions are shown in Figure~\ref{fig8}
by dotted and broken line histograms. They were important in
establishing the sensitivity of the results to the
presently poorly constrained value of $f$. Because of the large
number of the free parameters (5 bins in $Q(q)$) relative to
the number of independent data (7 or 8 bins in $A(a)$ for
both BW samples),
the individual uncertainties for each bin of $Q(q)$ 
were very large, in fact much larger than the Poissonian 
uncertainties. Still, it did not prevent us from iterating
the random search to a unique and stable solution for each value
of $f$. All solutions turned out to be very similar for all three
values of the fill-out parameter, in spite of the very
poorly constrained search. This leads us to conclude
that, irrespectively of the assumed value of $f$, the 
mass-ratio distribution appears to steeply rise 
for very small values of $q$; at the present level of accuracy,
the matter of the degree-of-contact is of secondary importance. For 
that reason, the subsequent analysis will consider only
the case of $f=0.25$.

\subsubsection{Power law approximations}

The next assumption is that the mass-ratio distribution can be
represented by a power law. Here we have a choice of expression,
either $Q_a(q) = a_0 (1-q)^{a_1}$, with $a_1 > 0$ or 
$Q_b(q) = b_0 q^{b_1}$, with $b_1 < 0$. 
The first form is a bit more convenient because $a_0$
multiplies a factor which is confined between 0 and 1, so that one
has a clean separation of the shape dependence, controlled by
$a_1$, from the normalization, controlled by $a_0$. The latter form
is preferable for comparison with theory which usually involves
straight power-law expressions in $q$. 
The function $Q_b$ tends to infinity for
$q \rightarrow 0$; however, there exists a low limit to $q$
that prevents the divergence, as we discuss in Section~\ref{min_q}.
Both forms involve only two parameters. 
We found that we cannot, at the present time, generalize them  
by addition of an additive parameter. 
Tests of the significance of such a third
parameter indicate an insignificant decrease of $\chi^2$ so that
two-parameter fits must currently suffice. 
An absence of the additive term in $Q(q)$ agrees with
the expectation that $Q(q) \rightarrow 0$ for $q \rightarrow 1$,
an effect which is caused by a thermal instability at $q=1$ 
\citep{luc76,fla76}; we discuss this further in Section~\ref{evol}.
The calculations of the predicted $A(a)$ were
made with 10 bins of $\Delta q =0.1$ with the distributions
$A_q(a)$ as given in Table~\ref{tab1}.

\placetable{tab4}

The solutions for $Q_a$ and $Q_b$ are given in Table~\ref{tab4}.
In terms of the quality of fit $\chi^2$, the solutions 
for $Q_a$ are slightly better than those for $Q_b$. 
Figure~\ref{fig9} shows the $\chi^2$ contours corresponding to the
one-sigma errors of both parameters for both solutions
based on admissible range of $\Delta \chi^2 = 2.3$ above 
$\chi^2_{min}$ for the 68 percent significance level. Because of the
very small number of data in the observational $A(a)$ and of the
correlation between the multiplicative and power parameters,  
the errors of the parameters are large. At this point,
we are unable to decide which power law is the correct one.

\placefigure{fig9}

\placefigure{fig10}

Figure~\ref{fig10} illustrates the $Q(q)$ solutions for both power laws
(left panels) and the implied amplitude distributions (right
panels). The continuous and broken lines show the best fitting 
$Q_a(q)$ and $Q_b(q)$ distributions and the resulting $A(a)$
distributions. Qualitatively, the two forms of $Q(q)$ 
appear to be similar in the range $0.12 \le q \le 1$ 
which maps into $A(a>0.3)$. However, the two power laws  
do differ at the quantitative level when specific predictions
for populations of individual bins are compared. For example,
we can follow \citet{van78}, who compared the
ratios of the populations in the bins $0.1 < q < 0.2$ 
and $0.8 < q < 0.9$ (see below in Section~\ref{evol}).
We find that the better fitting law $Q_a$ predicts the 
ratio of populations of these bins of $20 \times 10^3$
while $Q_b$ predicts the ratio of 37.
Unfortunately, with only 50 systems spread over 7 amplitude bins for
BW$_3$ and with 120 systems spread over 8 amplitude bins
for BW$_5$, the distributions $A(a)$ cannot 
constrain the results any better.

For further considerations we will simplify the results to
$Q_a(q) \propto (1-q)^{a_1}$, with $a_1 = 6 \pm 2$, 
and $Q_b(q) \propto q^{b_1}$, with $b_1 = -2 \pm 0.5$. 
This simplification is justified in view of (1)~the large
parameter errors for both power laws so that our
results are only very preliminary, (2)~the difference in the
results for the BW$_3$ and BW$_5$ samples,
and (3)~our preference for the BW$_3$ sample which is a 
very small, yet is more rigorously defined.

\subsection{Expected discovery selection at low amplitudes}

Having the predictions of the amplitude distributions which best
fit the amplitude range $a>0.3$, we can check how many systems
would be predicted over the whole range of amplitudes, including
those below this amplitude limit. These estimates
depend very strongly on the shape of $Q(q)$ for small
mass ratios. We give the predictions for the power-law
approximations of $Q(q)$ in Table~\ref{tab4} in the
line $\Sigma A_{pred}$. The numbers are of all systems expected
over the whole range of amplitudes. For the OGLE sample, we can
compare them with the actual numbers, including the systems with
small amplitudes below $a = 0.3$. 
The number of observed systems is 98 and 238, 
for BW$_3$ and BE$_5$, respectively.
For the $Q_a(q)$ distribution, the ratio
$N_{pred}/N_{obs} = 3.1$ is identical for both BW samples, but
for the $Q_b(q)$ distribution, the ratio is
8.8 and 12.9. Obviously, to a large degree, these estimates 
measure the amount of divergence of $A_{pred}(a)$
for $a<0.3$. They cannot be used to address the
important issue of the conversion of the apparent frequency 
to actual (spatial) frequency of contact binaries. There is
one effect which prevents the conversion factor from becoming 
uncomfortably high. It is the low limit on the mass ratio
which is described in the next section.

\section{THE MINIMUM MASS RATIO}
\label{min_q}

\citet{web76} pointed out that stability of a contact binary
is com\-pro\-mised by a re\-distribu\-tion of angular 
momentum for very small values of the mass ratio. 
For a mass ratio smaller than a threshold value
$q_{min}$, the system will find it easier to store its
angular momentum in one star rather than in an extreme mass-ratio
binary so it will quickly (in a dynamical time scale comparable
to one orbital revolution) merge into a single, rapidly rotating
star. \citet{ras95} re-analyzed this tidal instability
process and lifted the expected number from the very small value 
suggested by \citet{web76} to 
$q_{min} \simeq 0.09$. The exact location of the limit may
depend on the evolutionary state of the stars. This can
in fact explain the existence of such a well-known system
as AW~UMa with $q = 0.075$.

The distributions $A_q(a)$ sampled at $\Delta q =0.1$, 
as in Table~\ref{tab1} and Figure~\ref{fig3}, 
are obviously useless in predicting $A_{pred}$ 
in the presence of the cut-off at 
$0.07 \le q_{min} \le 0.09$. However, we can use
these distributions to obtain a preliminary (upper limit) 
estimate on $A_{pred}$ by simply setting $Q(q)=0$
for the first bin $0 < q < 0.1$. The predicted number of all systems 
is then substantially reduced, for the $Q_a$ law from
301 and 732 systems to 143 and 350
and, for $Q_b$ law from
850 and 3082 to 147 and 359, for BW$_3$ and BW$_5$,
respectively. Thus, the ratio $N_{pred}/N_{obs}$, which
was as large as 3 to 12 with the full $Q(q)$ distributions
discussed in the previous section, is reduced to more
acceptable levels: $N_{pred}/N_{obs} = 1.46$ and 1.47 for $Q_a(q)$
and $N_{pred}/N_{obs} = 1.50$ and 1.51 for $Q_b(q)$. Similarity
of these numbers at about the level of about 1.5
attests to the fact that the power-laws
$Q_a$ and $Q_b$ are both equally successful 
when only the amplitudes $a>0.3$ and the
mass ratios $q>0.1$ are considered.

The above estimate is approximate because the cut-off is
almost certainly below $q_{min} = 0.1$. To reproduce the
shape of $Q(q)$ below this point for use in Equation~\ref{eq2},
we must resort to the original fine grid of $A_q(a)$ calculated with  
small bins of $\Delta q =0.01$. The results for the specific case 
of the BW$_3$ sample and for both power laws are shown in 
Figure~\ref{fig11}. Note the particularly 
strong influence of the cut-off at $q_{min}$ on the $Q_b$ results
when the very steep increase in $A(a)$, due to the divergence 
of $Q_b$ for $q \rightarrow 0$, is avoided.

One can calculate integrals of the curves in Figure~\ref{fig11}
and take their ratio. 
Values of the ratio $N_{pred}/N_{obs}$ for $q_{min}=0.10$, 
0.08 and 0.06 (and for an imaginary extension down to $q=0.01$) 
are given in Table~\ref{tab5}. $N_{pred}/N_{obs}$ 
is the correction factor which can be used in converting the
apparent (inclination uncorrected) frequency of occurrence of W~UMa
systems -- as derived on the basis of the BW$_3$ sample -- 
into the true spatial frequency of occurrence.
The uncertainty with the value of $q_{min}$ prevents us from establishing
this factor to any better than about 1.5 to 2.0. Thus,
the currently best estimate of the inclination-uncorrected
frequency of contact binaries in the old disk population of
about 1/130 relative to FGK dwarfs \citep{ogle-cl} would translate
into the spatial frequency of about  1/80 -- 1/65. 

\placefigure{fig11}

\placetable{tab5}

The last two columns of Table~\ref{tab5} give the predicted
ratio of the total number of systems to those detectable by surveys
fully complete for amplitudes $a > 0.1$, for both power laws. 
The missed fraction for
such surveys would be very similar, with the correction factor
at the level of 1.4 to 1.5 times. The closeness of
the estimates of $N_{pred}/N_{obs}$ for the BW$_3$ sample 
(about $1.5-2.0$) and for a fully complete sample down 
to $a=0.1$ (about $1.4-1.5$) is due to
(1)~a flattening of $A(a)$ in the region $0.1 < a < 0.3$ caused
by the mass-ratio cut-off, and (2)~the fact that 
the OGLE sample, although probably not complete below $a=0.3$, 
contains a fair number of systems with $0.1 < a < 0.3$. 

The shape of a well defined amplitude distribution $A(a)$ down to 
$a=0.1$ is expected to sensitively reflect the location of $q_{min}$.
We can see in Figure~\ref{fig11} that a local maximum 
is expected to form in $A(a)$ around $a \simeq 0.2 - 0.25$.
This maximum is better defined for the $Q_b(q)$ power law
because the rise of the predicted $A(a)$ below $a=0.3$ is steeper
for this law so that the effects of the cut-off in $q$ are stronger.
We actually see a maximum in the observed $A(a)$ for the OGLE
sample exactly in this interval, but we suspect that this feature
is simply due to the detection selection effect setting in
for $a<0.3$. To be sure of the existence of the local maximum, 
we should see indications of a small minimum beyond the peak
and of the further increase in $A(a)$ for $a < 0.15-0.2$.
As we have said above, an extension of completeness down to
$a=0.1$ will still leave some 40 to 50 percent of all systems
below the detection level.

\section{THE MASS-RATIO DISTRIBUTION 
IN THE CONTEXT OF CONTACT BINARY EVOLUTION}
\label{evol}

The mass-ratio distribution for W~UMa-type systems
and the evolution of this distribution over time are related to 
entirely different processes than those of star formation 
producing an almost flat $Q(q)$ for detached binaries 
\citep{maz92}. Contact binaries have the freedom of exchanging
mass between components through a complex interplay of
energy exchange, mass exchange and angular-momentum loss (AML) 
processes. The first major re-structuring takes place at
the moment when the contact system forms from two 
detached components; from that point, 
further, more gradual changes in the mass
distribution are expected as the system evolves over time.

Following the pioneering works of \citet{luc76} and \citet{fla76} who
showed that contact systems are inherently thermally unstable and
will evolve away from $q=1$ to small mass ratios, 
several theoretical models explored in detail
the thermal-relaxation oscillations and ways of preventing them,
either through nuclear evolution or through AML, 
or perhaps combination of both processes 
\citep{RE77,rah81,rah82,rah83}. At this moment,
the unified scenario of the contact binary formation 
and evolution presented by \citet{vil82}, appears to be still 
valid. Among its unexplained features, the most mysterious 
remains a coupling and/or feedback process between the degree
of contact and the amount of AML which prevents rapid coalescence
on one hand and disruption of contact (a semi-detached phase)
on the other hand. Our understanding of these processes
crucially depends on the estimates of the relevant evolutionary
time scales which can be estimated from the statistics of $Q(q)$.

The discussion of $Q(q)$ in \citet{vil81} revolved around the 
(then) only available observational derivation
of \citet{van78}, based on 
very meager, combined photometric and radial-velocity data. 
These results most probably contained strong
discovery and observer preference selection effects, 
in spite of heroic attempts to estimate their size. 
This distribution was much less steep than the power laws 
derived in the current paper: \citet{van78}
estimated that his distribution implies ten times more systems
in the $0.1 < q <0.2$ bin than in the $0.8 < q < 0.9$ bin. 
Our power laws lead to a much larger disparity in the
population of these bins: The ratio predicted by
$Q_a \propto (1-q)^6$ is $20 \times 10^3$ times (there would be
almost no large mass-ratio systems), while $Q_b \propto q^{-2}$ 
predicts the ratio of
37 times. While the large difference in the predictions
will eventually help in selecting the correct shape for $Q(q)$,
we are not at present in a position to prefer one power
law over the other as both give very similar fits to the
observed $A(a)$ for $a>0.3$.

As discussed by \citet{vil81} (see Figure~4 in this paper), 
the mass-ratio evolution in contact is driven mainly by the 
less-massive component and its thermal (Kelvin--Helmholtz)
time scale. When the evolution reaches a steady state condition,
the number of systems in a particular evolutionary stage
should scale as $N \propto \tau_{sec} \propto M_{sec}^{1-\beta}$, where
$\beta$ is the exponent in the mass--luminosity relation,
$L \propto M^\beta$. For the lower main sequence, $\beta \simeq 4.5$,
so that $\tau_{sec} \propto M_{sec}^{-3.5}$. Since
$M_{sec} = M_{tot}\,q/(1+q)$, for small values of $q$, we can expect 
$Q(q) \propto q^{-3.5}$. Thus, 
as the secondary components become less massive, their 
evolutionary time scale becomes progressively longer
resulting in a pile up of contact systems 
at low mass ratios. This pile-up is
limited by the tidal instability at $q_{min}$,
as discussed in Section~\ref{min_q}. 

The theoretical expectations described above very 
well agree with our results which we present in a schematic
form in Figure~\ref{fig12}. We note that location of the
tidal instability at $q_{min} \simeq 0.07-0.1$
is a relatively new development \citep{ras95} and could not 
be included in the
general discussion of \citet{vil81}. Its presence is actually 
crucial in preventing the problem of embarrassingly too many
small mass-ratio systems, if the $Q(q) \propto q^{-3.5}$ distribution
were to continue below $q \simeq 0.07-0.1$.

\placefigure{fig12}

Finally, a cautionary note: If the evolution were really to slow
down as the mass of the secondary component decreases, i.e.\
as $\tau_{sec} \propto M_{sec}^{-3.5}$ then, for very small
$M_{sec}$, it would take longer than the Hubble time.
Assuming the thermal time scale for the Sun, $\tau_\odot \simeq
3 \times 10^7$ years, then for $M_{sec} = 0.1\,M_\odot$, 
$\tau_{sec} \simeq 10^{11}$ years. This would lead to a very
inefficient, practically insignificant formation of 
single, rapidly-rotating stars from contact binaries.
But -- more likely -- the nuclear or AML evolution of
primary components, with the associated shorter
time-scales, will become more important
first. Thus, we have no idea about the rate of production of single
stars at the cut-off at $q_{min}$, but it may be not as low
as the thermal evolution of secondary components would imply.

\section{CONCUSIONS}
\label{concl}

This paper discusses the expected amplitude
distribution $A(a)$ for contact binary stars. The strong dependence of 
$A(a)$ on the mass-ratio distribution, $Q(q)$, has been shown 
to be useful for shedding light on the latter distribution which
has a considerable astrophysical significance. We attempted to
simplify the details of the approach and to concentrate on the
main properties of both distributions. In particular,
while the results
do depend on the degree-of-contact parameter, $f$, the
dependence is weak and -- for simplicity -- most of the discussion
has been presented for the most likely value of $f=0.25$. 

The main limitation of the paper are problems created by
presence of ``third light'' in photometry of contact binaries.
Both, presence of unresolved visual companions and of blending 
in crowded areas such as Baade's Window, are expected to produce
distorted amplitude distributions with a diminished
representation of systems with large amplitudes. 
The degree of such a distortion is difficult to quantify, 
primarily because the unknown frequency of occurrence
of companions to contact systems; this frequency may be
different than for wider binaries. The data for the 7.5 magnitude-limit
sky-field sample indicate that the main conclusions of the
paper are valid even after accounting for presence of
close companions.

The main results of the paper are summarized below with references
to Figure~\ref{fig12}.
\begin{enumerate}
\item The two distributions, of the mass ratio, $Q(q)$,
and of the photometric variability amplitude, $A(a)$,
are very closely inter-related. Since the $Q(q)$ distribution
is expected to
contain a record of the contact binary evolution, studies
of $A(a)$ can help in resolving the still poorly understood
details of time scales of formation and evolution of contact systems.
\item The amplitude distribution $A(a)$ is expected to rise
for small amplitudes almost irrespectively of
the shape of $Q(q)$. This rise is expected for a flat
distribution, $Q(q)=const$, or even for any of imaginary
``monochromatic'' distributions with all systems having just
one mass ratio, $Q(q) = \delta(q-q_0)$. 
The rise becomes even stronger
if $Q(q)$ steeply increases for small $q$, as it appears to be
the case. 
\item The increase of $A(a)$ for $a \rightarrow 0$ 
continues to zero amplitude and leads to a convergence to a 
constant (mass-ratio dependent) value: 
$A_q(a \rightarrow 0) \rightarrow C(q)$.
\item Two samples of contact binaries have been considered:
The sample of bright systems to 7.5 magnitude and the sample
of systems discovered in the Baade's Window by the OGLE project. While
the former appears to be complete to $a \simeq 0.1$ 
and has been corrected for the presence of 
currently known companions, it is
too small for derivation of $Q(q)$ from $A(a)$. The latter sample
is marginally sufficient in size (98 or 238 systems depending on the
spatial depth) and gives a moderately 
well-defined amplitude distribution,
but is only complete for $a>0.3$ and is
certainly subject to the influence of blending problems
which tend to depress the large amplitude end of $A(a)$.
The conclusions below are preliminary on account of the
neglected photometric blending for the OGLE sample.
\item Thanks to the non-linearity of the relation between $A$ and $Q$,
an amplitude distribution complete for
$a>0.3$, maps into $Q(q)$ within $0.12 \le q \le 1$,
so that the accessible range of $q$ in $Q(q)$ is respectable. 
Figure~\ref{fig12} shows our best power-law estimates of $Q(q)$. 
\item The $Q(q)$ distribution derived
from the OGLE distribution $A(a)$ for 
the interval $0.12 \le q \le 1$
climbs very steeply for $q \rightarrow 0$; it can be
approximated by $\propto (1-q)^6$ or $\propto q^{-2}$. The
values of the exponents in both expressions are very preliminary,
not only because of large statistical errors but -- more
importantly -- because of the distortions to $A(a)$
introduced by the blending effects.
\item The steep increase of $Q(q)$ is expected to 
be abruptly terminated at $q_{min} \simeq  
0.07 - 0.1$ by the process of tidal instability. This alleviates
the danger of huge numbers of very small amplitude systems
which would be hiding below the detection thresholds,
if the approximate power-law relationships were
to continue to $q \rightarrow 0$.
\item The most common contact systems in the interval
between the cut-off at $q_{min} \simeq 0.07 - 0.10$ 
and the steep power-law drop at
$q \simeq 0.3 - 0.4$ are expected to 
generate a local maximum in the amplitude distribution in the vicinity
of $a \simeq 0.20 - 0.25$. The exact location of this maximum
and the rate of increase of $A(a)$ for $a \rightarrow 0$
will help to establish the value of $q_{min}$ which is
currently poorly established. 
\item It is expected that future
well-determined amplitude distributions, good down at least to
$a \simeq 0.1$ and with fully characterized blending
will define the exact shape of $Q(q)$
in the vicinity of the cutoff at $q_{min}$. The current sky-field
sample to $V = 7.5$ contains too few systems for a good
definition of $A(a)$ at small amplitudes; a deeper
sample is needed. 
A complete sky-field sample has a potential of a better control over
the problem of unresolved companions than the OGLE sample
because of the accessibility to spectroscopic studies.
\item The previous analysis of the OGLE sample led to an
estimate of the inclination uncorrected frequency of 
contact binaries of about 1/130 relative to FGK dwarfs.
The OGLE sample contains contact systems with the
smallest amplitudes of about 0.1 mag.\ and appears to be
fully complete for $a>0.3$ mag. At present, we estimate that
a correction factor to convert the OGLE apparent frequency into 
the {\it spatial\/} frequency is
about 1.5 to 2.0, but the exact value sensitively depends
on the value of $q_{min}$. Thus, the inclination-corrected
spatial frequency is one contact binary per 1/80 to 1/65
Disk Population FGK dwarfs.
\end{enumerate}

\acknowledgments

I would like to thank Stefan Mochnacki for useful discussions
about various subjects related to statistics of contact binary stars,
to Greg Stachowski for 
careful reading of the manuscript and for linguistic corrections
and to Janusz Ka{\l}u\.{z}ny who -- acting as a referee --
provided several useful suggestions for improvement of the
final manuscript.

The author acknowledges with gratitude the fact that this
paper utilizes the data obtained and made available for
public access by the OGLE project.

Support of the Natural Sciences and Engineering Council of Canada
is acknowledged with gratitude.

\clearpage

\noindent
Captions to figures:

\figcaption[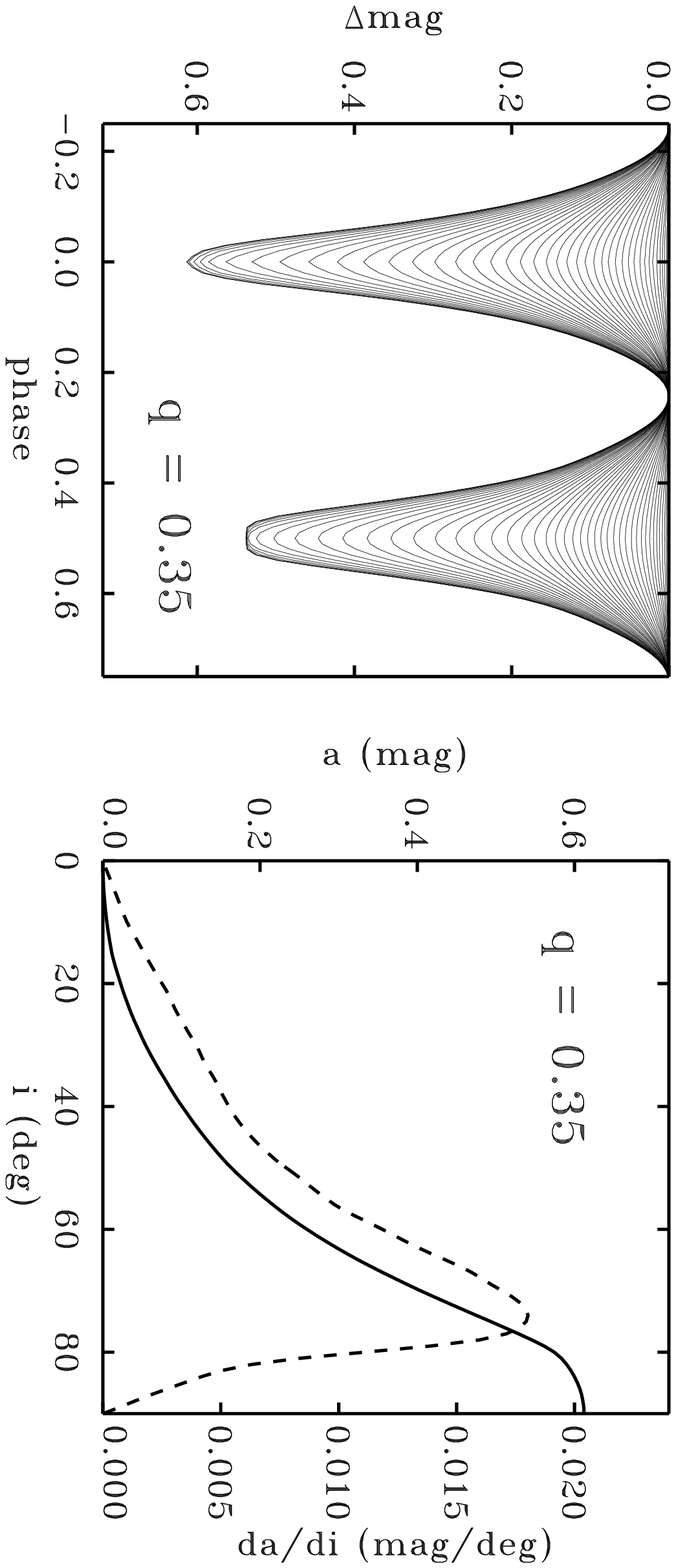]{The left panel shows 45 light curves with
the orbital inclination angle $i$ varied in steps of 2 degrees between 
0 and 90 degrees for a case of $q=0.35$ and $f=0.25$. 
The right panel shows the corresponding change of the
amplitude, $a=a(i)$ (solid line) as well as its derivative,
$da/di$ (broken line and right vertical scale).
The amplitudes in this and subsequent figures are expressed
in $V$-band magnitudes.
\label{fig1}}

\figcaption[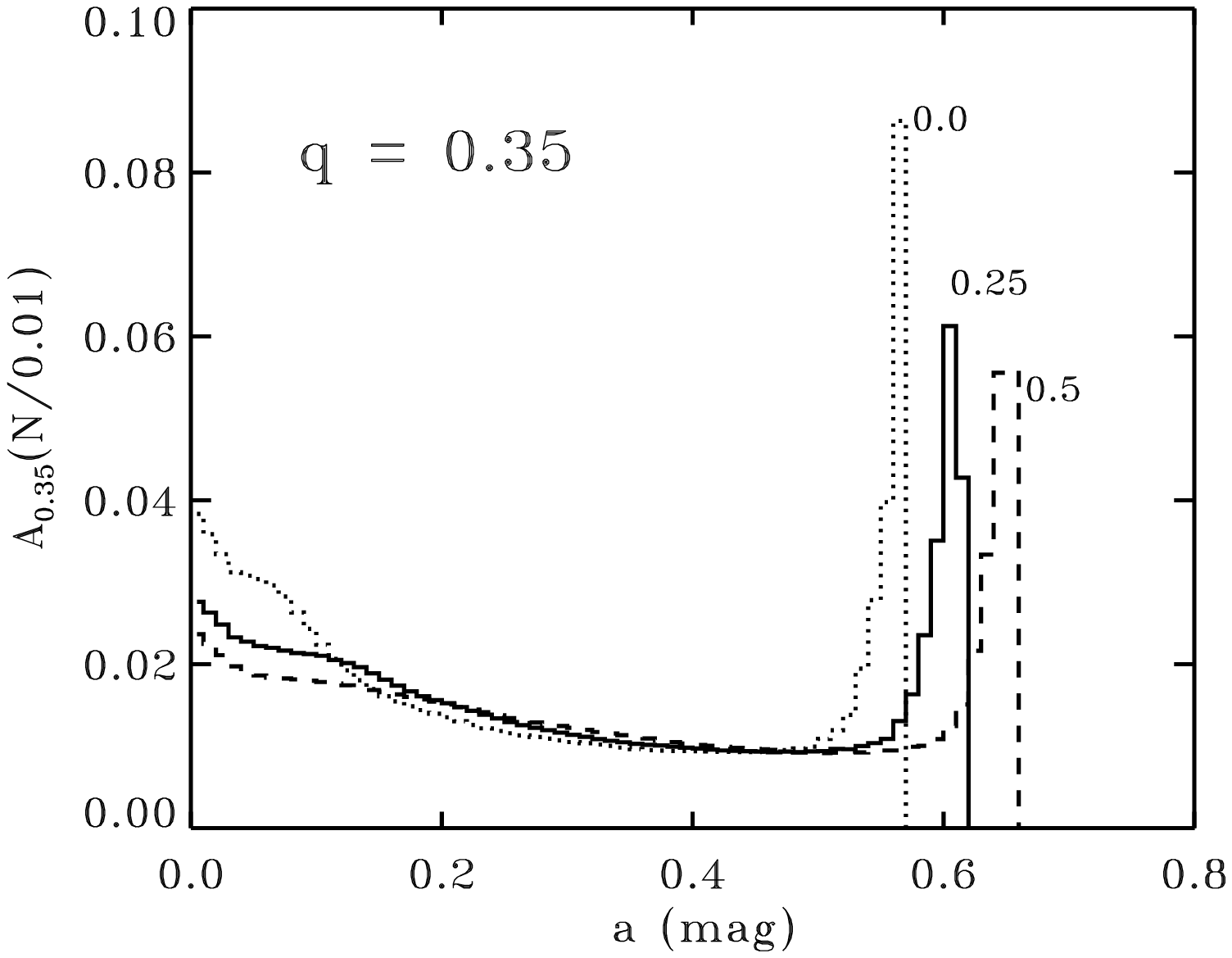]{The amplitude distribution $A(a)$ for the
case of $q=0.35$, for three values of the degree-of-contact
parameter $f$, as labeled in the figure. The distribution
was obtained for a narrow range $\Delta q = 0.01$ so that the
maximum of $A(a)$ corresponding to total eclipses is very narrow.
The results on $A(a)$ computed with $\Delta a = 0.01$, 
such as shown in this figure, suffer from small-scale inaccuracies
for $a < 0.15$, mostly because of the  
imperfections in the light-curve synthesis 
model for very low inclination angles. Distributions for wider
intervals of $\Delta q$, sampled
into larger bins $\Delta a$ avoid these inaccuracies.
\label{fig2}}

\figcaption[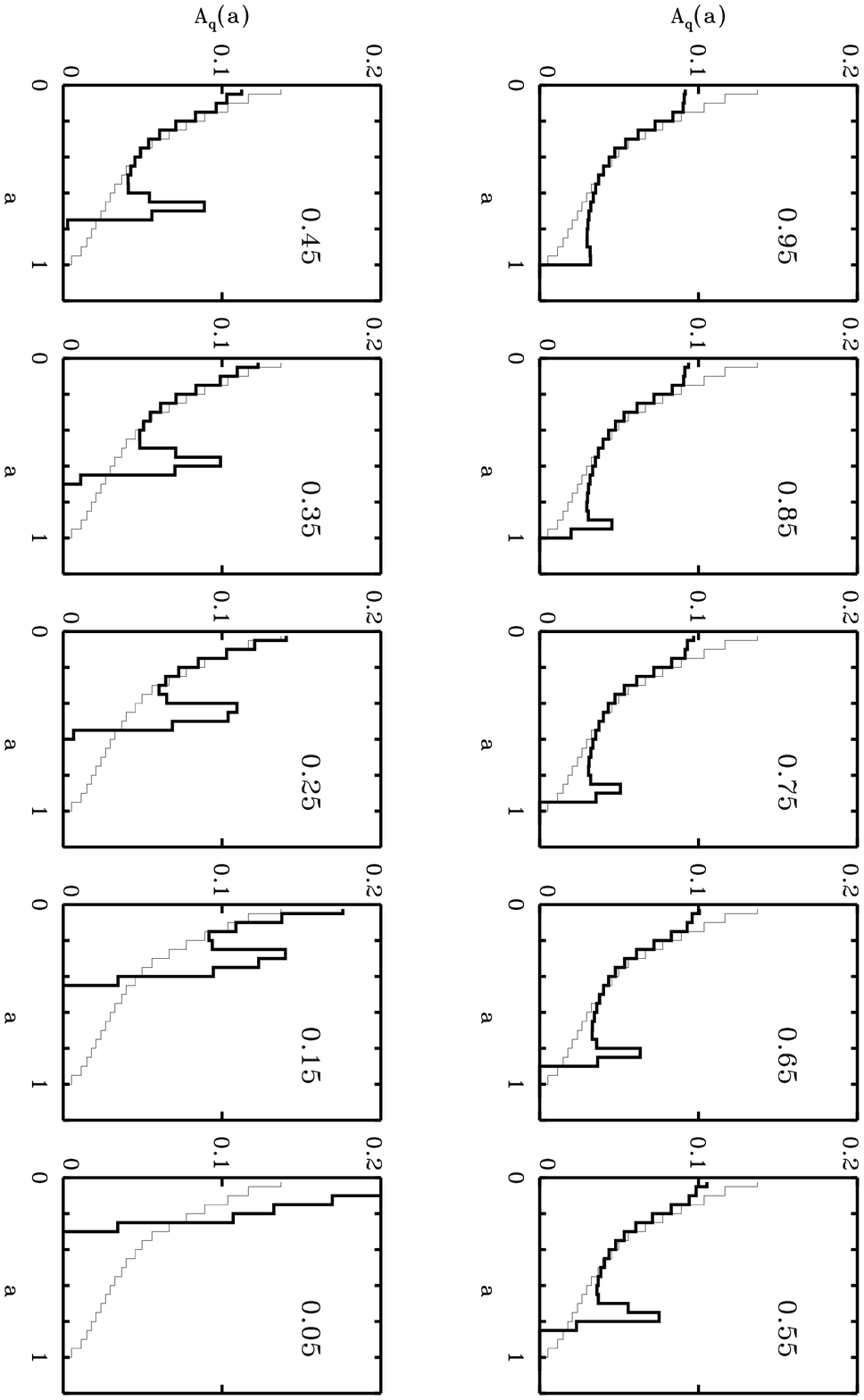]{The amplitude distributions $A_q(a)$ for
$f=0.25$ calculated in bins $\Delta a = 0.05$ and in intervals
$\Delta q = 0.1$. The values of $q$ for the bin centers are
given in each panel. The thin line gives the normalized
summed distribution corresponding to a flat distribution
$Q(q) = const$.
\label{fig3}}

\figcaption[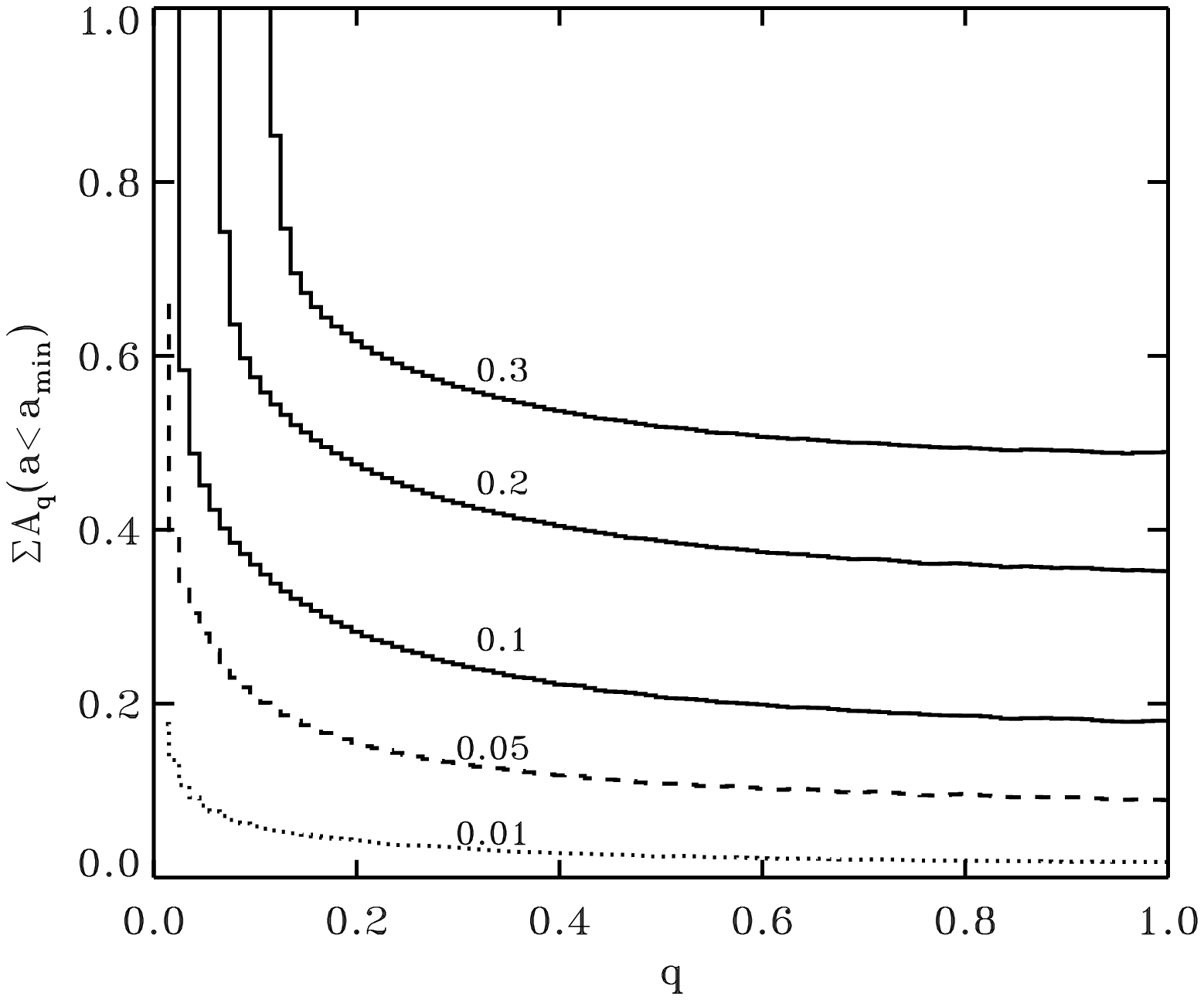]{The fraction of systems with amplitudes smaller
than a limiting amplitude, given as a label of the
line, is shown here as a function of the mass ratio $q$. The distributions
$A_q(a)$ used for this figure were computed in intervals of 
$\Delta q = 0.01$.
\label{fig4}}

\figcaption[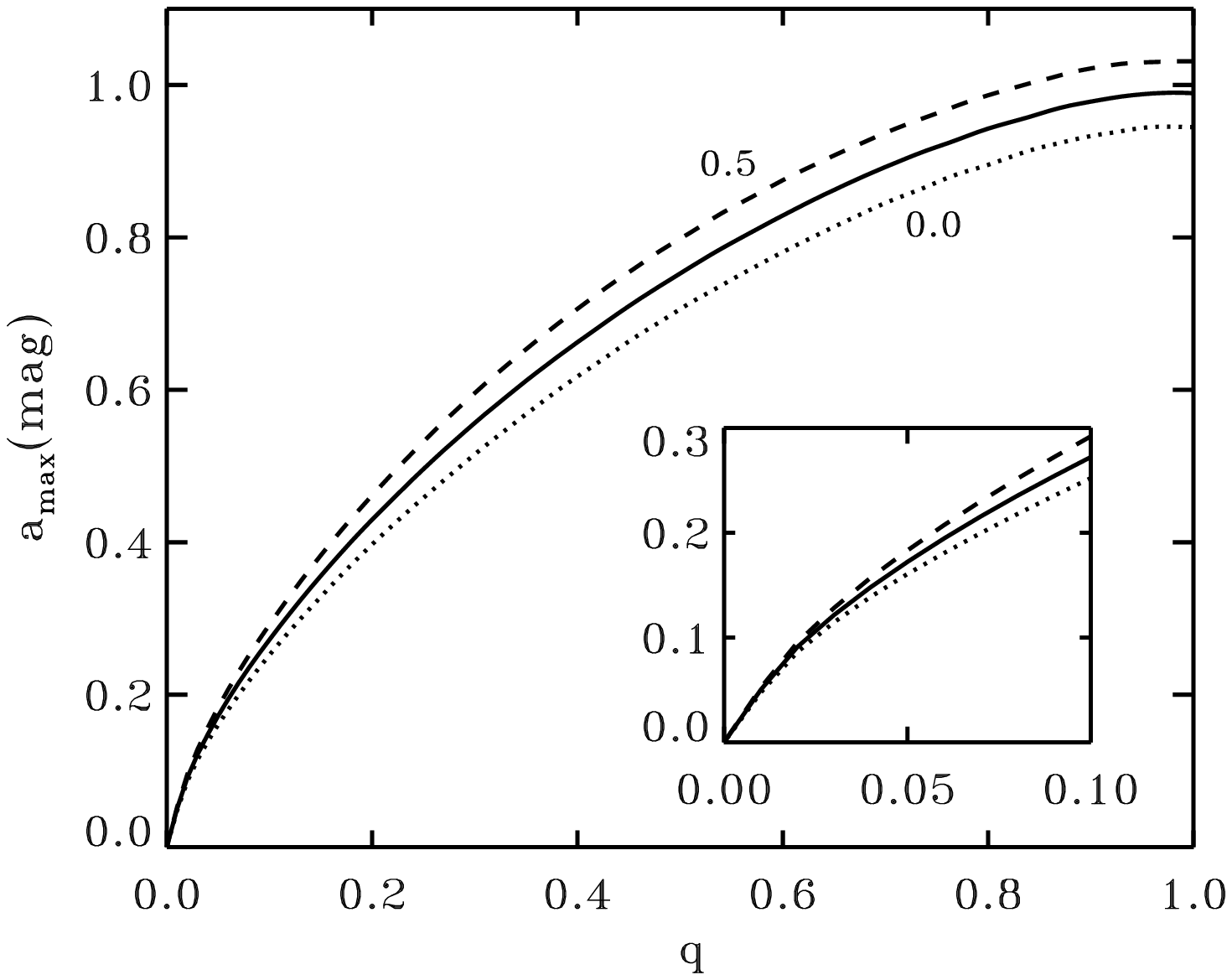]{The maximum amplitude $a_{max}$ as the function
of the mass ratio, $q$ for three values of the degree-of-contact parameter
$f$. The case of $f=0.25$ is shown by the continuous line and the cases
of $f=0$ (inner contact) and $f=0.5$ are shown by dotted and broken lines.
The insert shows the details for very small mass ratios. 
\label{fig5}}

\figcaption[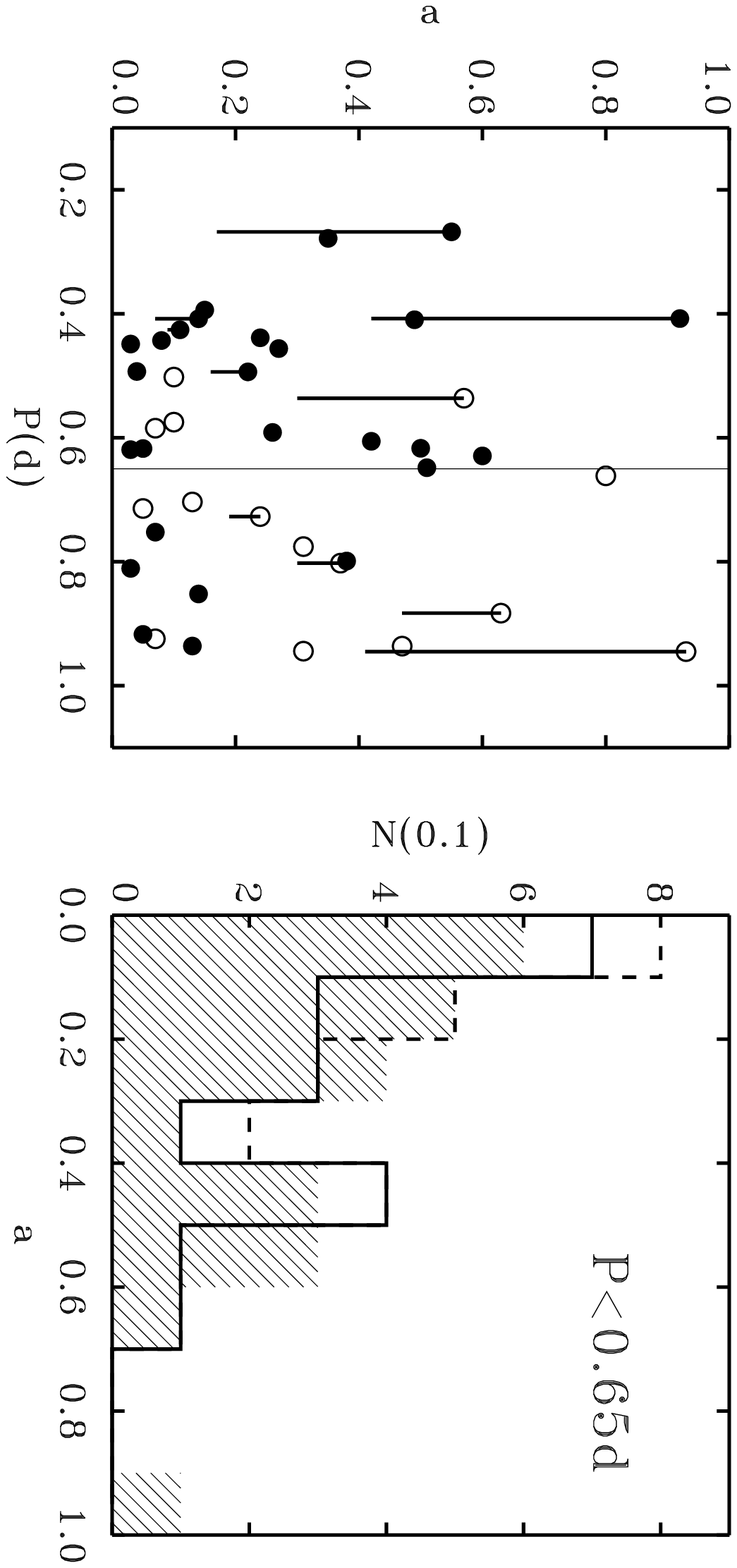]{The amplitudes of contact (EW -- filled circles), 
semi-detached or poor-thermal-contact systems (EB -- open circles) are 
shown as the function of the orbital period
for 41 binaries of the ``7.5 magnitude-limit'' 
sky-field sample in the left panel of the figure. 
This sample is most likely complete for amplitudes $a \ge 0.1$. 
The sample is not volume limited, so that intrinsically 
bright, long-period systems are preferentially included.
It is not always possible to unambiguously assign 
the class, EW or EB, but we note the dominance
of contact systems for periods shorter than 0.65 days, 
confirming was found for the OGLE volume-limited samples  
\citep{ogle-cl}. The vertical vectors show corrections to
amplitudes due to the presence of close companions. The 
combined amplitude distribution for 24 systems with periods shorter
than 0.65 days is shown in the right panel of the figure. The
shaded histogram shows the combined distribution of amplitudes
which have been corrected for the companions. The uncorrected
distributions are shown by line histograms, by the continuous line 
for 20 EW systems and by the broken line for the additional
4 short-period EB systems. 
\label{fig6}}

\figcaption[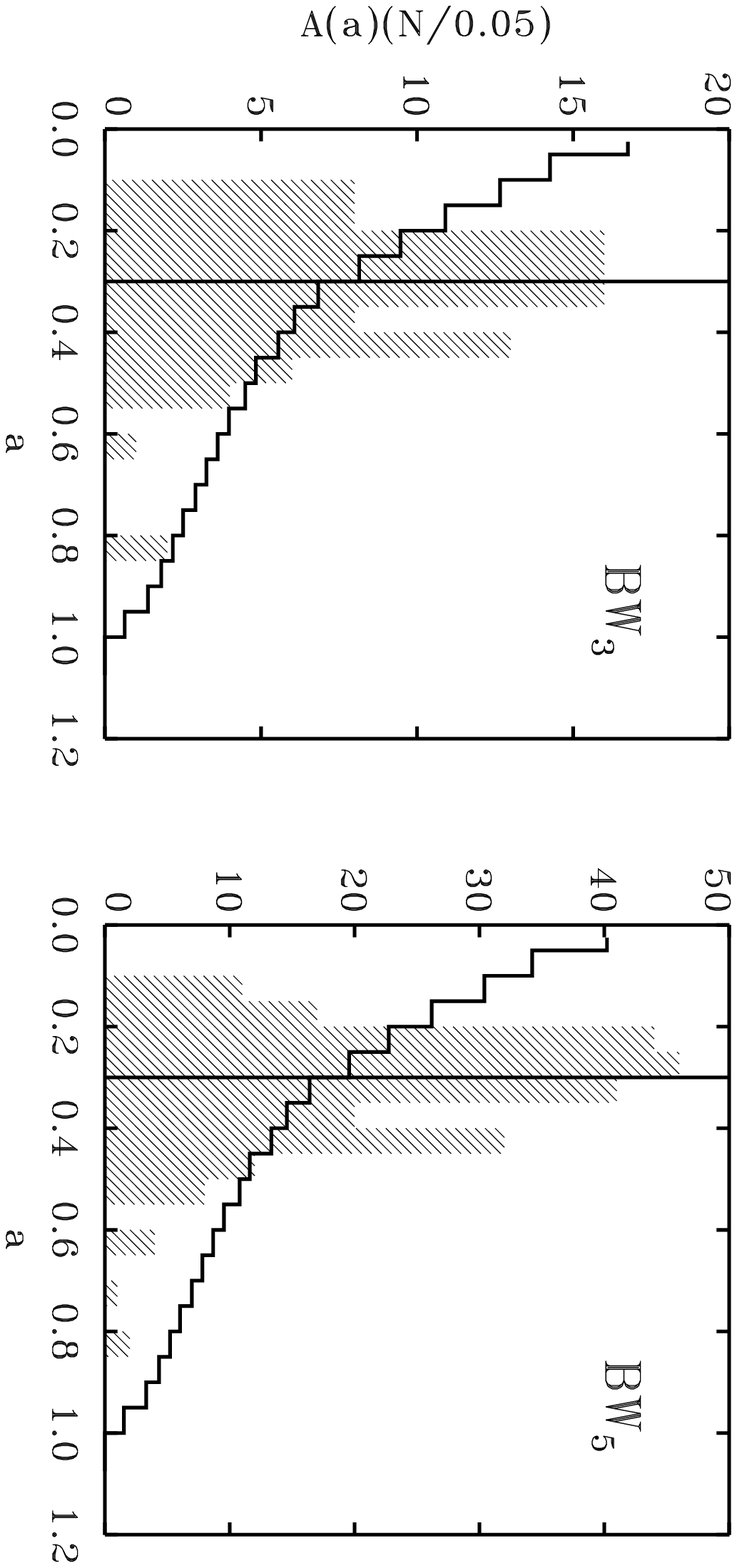]{The observed amplitude distributions for the
OGLE samples BW$_3$ and BW$_5$ are shown by shaded histograms. 
They are based on data obtained in the photometric $I$ band; thus, 
we commit a small inconsistency in this paper by comparing them
with the predictions for the $V$-band. The
predicted amplitude distribution for a flat distribution of the
mass ratio, $Q(q) = const$, is shown by a continuous line histogram.
We assume in this paper that the detection efficiency of the 
OGLE survey dropped below $a \simeq 0.3$ (vertical line). 
\label{fig7}}

\figcaption[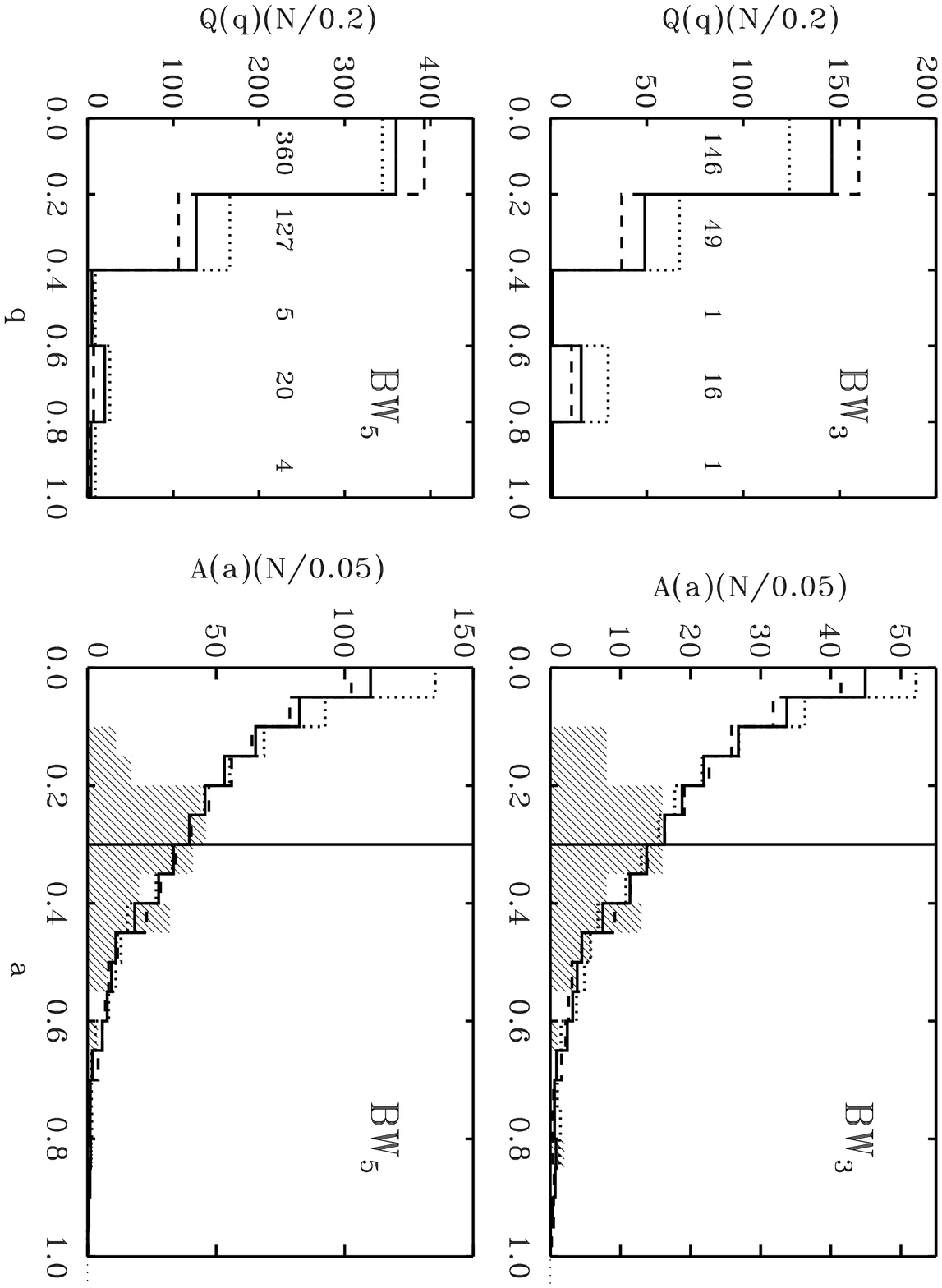]{The best fitting mass-ratio distributions
$Q(q)$, sampled as 5 independent values in intervals of 
$\Delta q = 0.2$, are shown in the two left panels of the figure. 
They correspond to the observed OGLE-sample BW$_3$ (upper
panels) and BW$_5$ (lower panels)
amplitude distributions, as shown in the right panels. The 
fits have been based only on the amplitude distributions for $a > 0.3$.
The solid lines show the solutions for $f=0.25$ while the dotted and
broken lines correspond to $f=0$ and $f=0.5$, respectively. 
The numbers above each bar give the solutions, in
numbers of systems, for $f=0.25$. The 
vertical lines in the $A(a)$ distributions delineate amplitudes
above which the OGLE data are almost certainly complete in terms of
discovery selection effects.  
\label{fig8}}

\figcaption[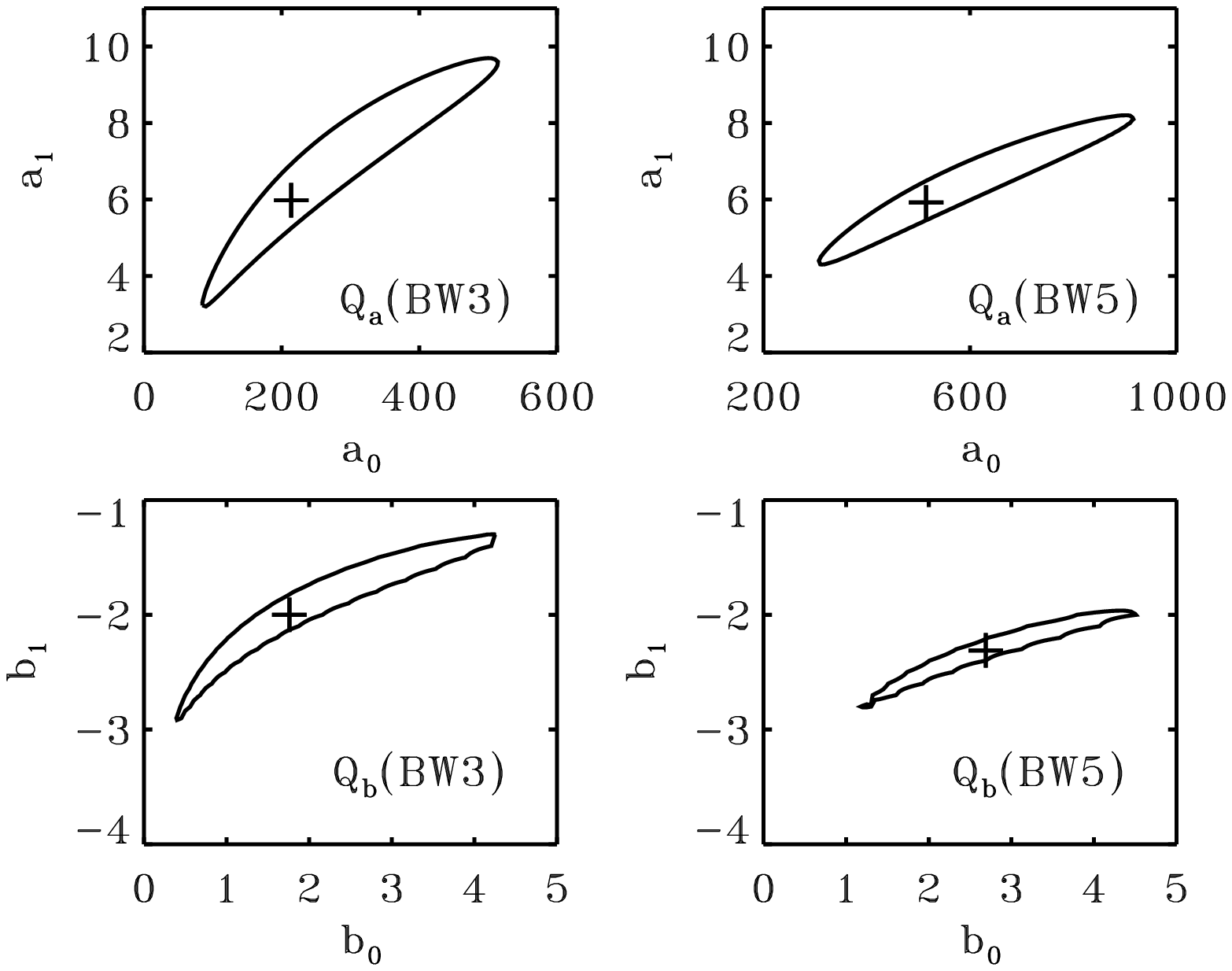]{The four panels show combinations of
the power-law parameters giving the best fits to the observed
amplitude distributions for the samples BW$_3$ and BW$_5$. The
power laws are: $Q_a(q)= a_0 (1-q)^{a_1}$ and 
$Q_b(q) = b_0 q^{-b_1}$. The minimum $\chi^2$ points are marked by
crosses while the contours give the 1-sigma levels $\Delta \chi^2 = 2.3$
above the minima, as appropriate for two-parameter fits. 
\label{fig9}}

\figcaption[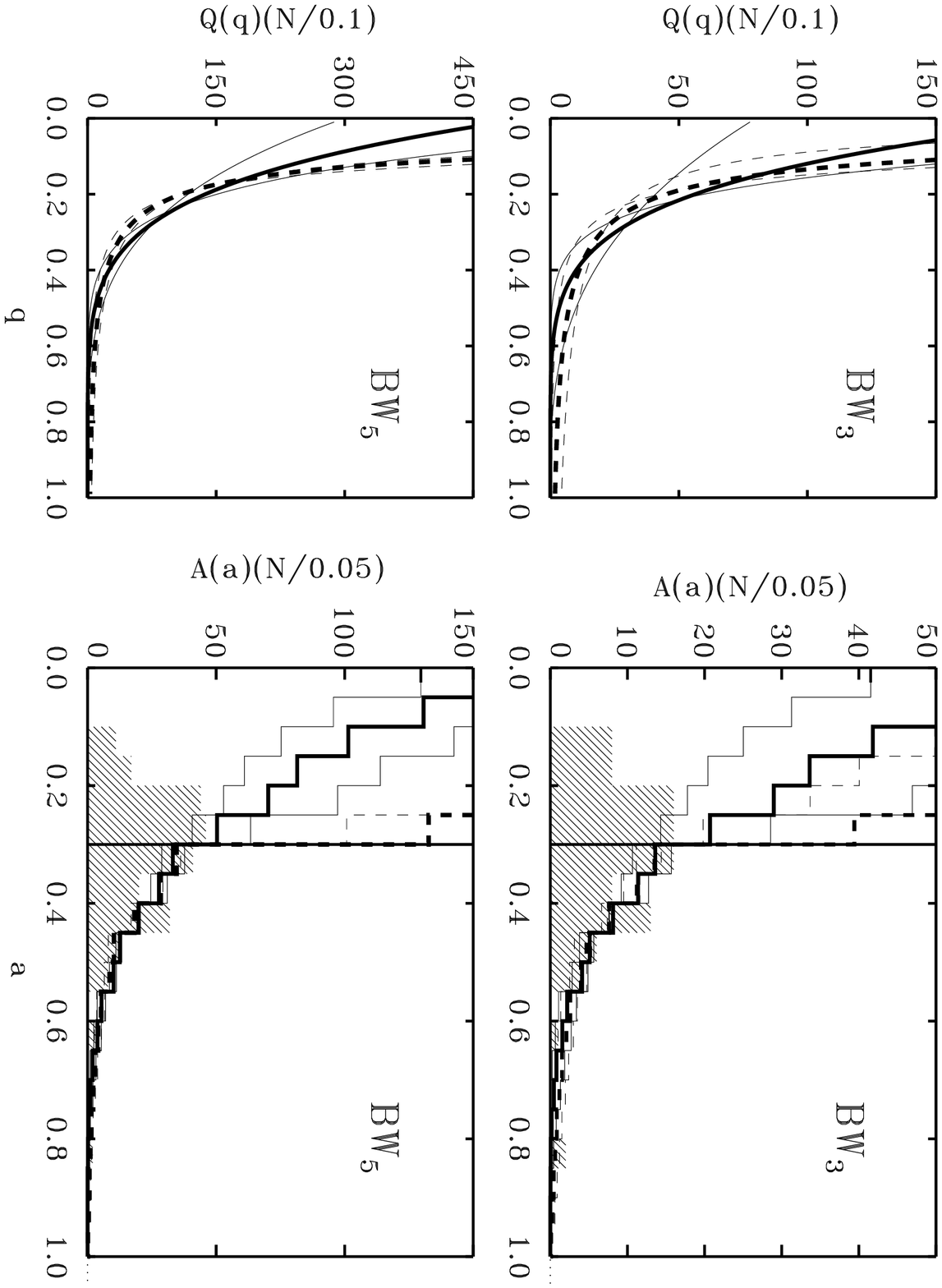]{This figure has a similar format to
Figure~\ref{fig8}. The thick continuous 
and broken lines show the best fitting 
$Q_a(q)$ and $Q_b(q)$ distributions and the resulting $A(a)$
distributions for both OGLE samples.
The corresponding thin lines give the predicted shapes
of $Q(q)$ and $A(a)$ assuming variation of the power law
parameters for the most extreme
ends of the $\chi^2$ correlation crescents in Figure~\ref{fig9}.
Note the very similar shape of all functions $A(a)$ for $a>0.3$
and the large divergence below this limiting amplitude.
\label{fig10}}

\figcaption[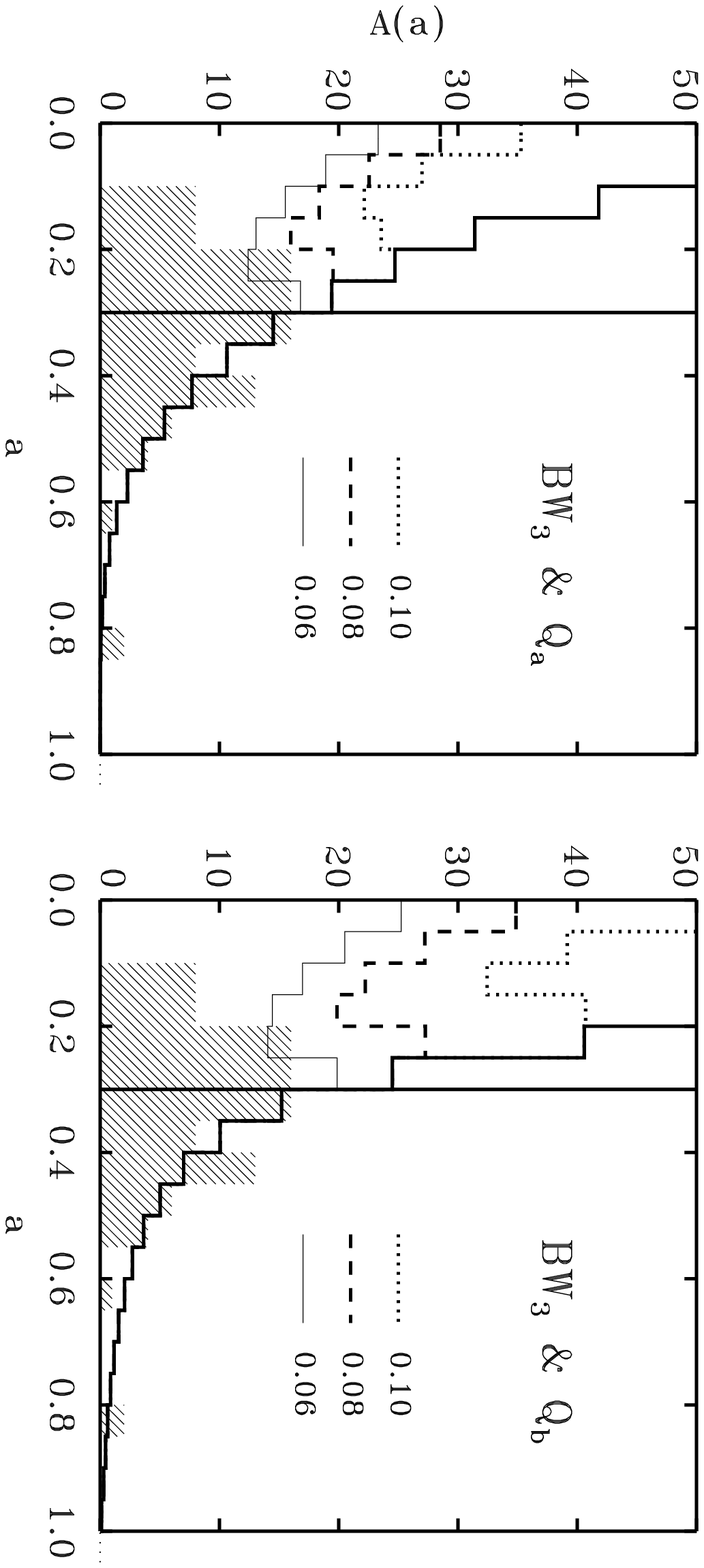]{The predicted amplitude distribution
is expected to be strongly modified at low amplitudes
by the low mass-ratio cut-off at $q_{min} \simeq 0.07 - 0.10$. 
The plots show the expected changes in $A(a)$ 
for $q_{min} = 0.1$, 0.08 and 0.06, for
the power-law distributions $Q_a(q) = a_0 (1-q)^{a_1}$
(left panel) and $Q_b(q) = b_0 q^{b1}$ (right panel), and
 for the best fit parameters as for 
the OGLE sample BW$_3$. The parameters of the fits are given in
Table~\ref{tab4}. Since the tidal instability causing
the mass-ratio cut-off must be present, the complete
statistical data will almost certainly show a local peak in the 
amplitude distribution at $a \simeq 0.2 - 0.25$ mag. The 
currently most trustworthy data of the OGLE sample appear
to be complete above $a>0.3$ (vertical line). 
\label{fig11}}

\figcaption[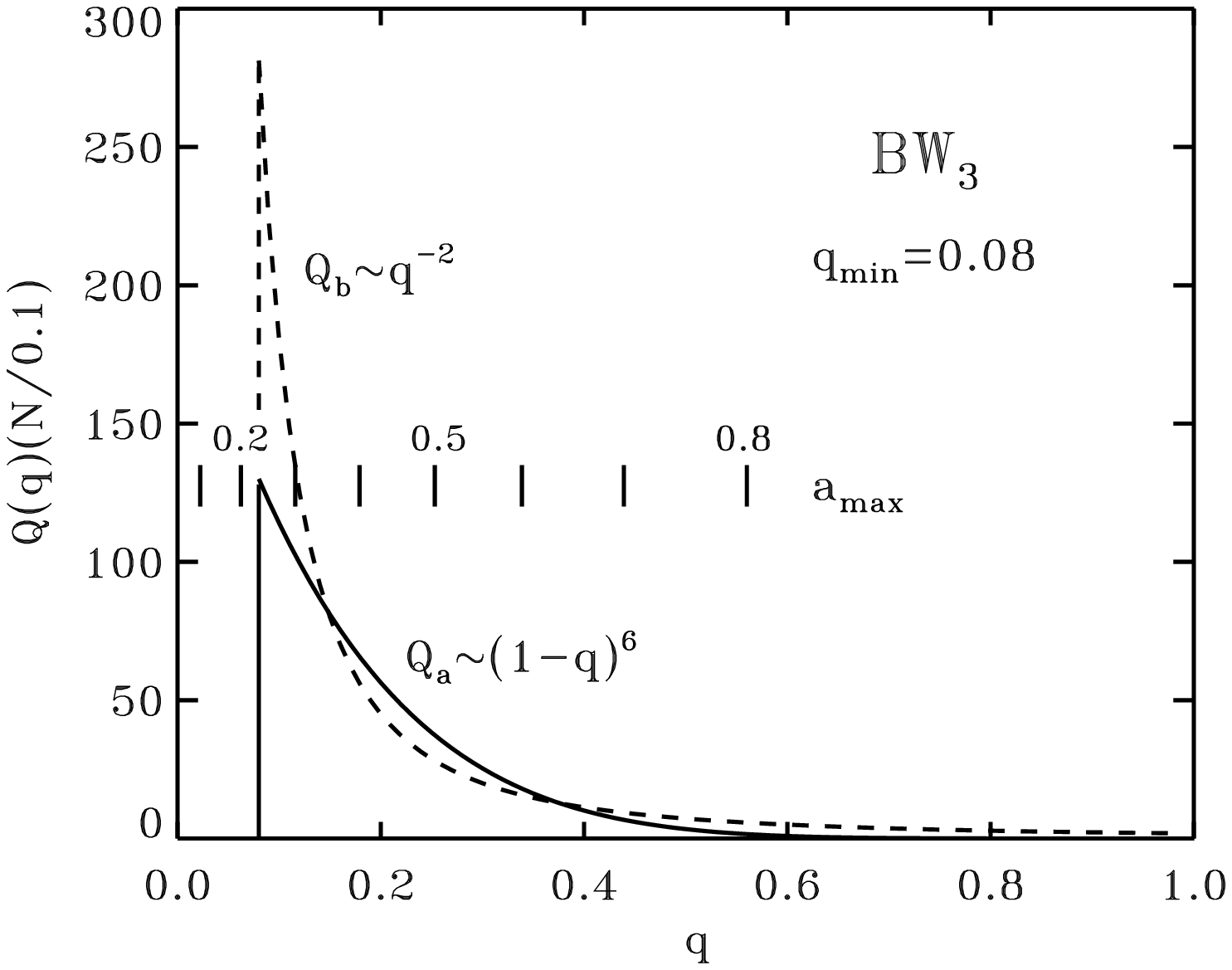]{The figure shows the most likely
shapes of the power-law distributions, as determined from
the OGLE sample amplitude distributions, which can be approximated
by $Q_a \simeq (1-q)^6$ or $Q_b \simeq q^{-2}$
(the values of parameters are given in Table~\ref{tab4}). 
The distributions
must experience a sharp cut-off at $q_{min} \simeq 0.07-0.10$. 
In the middle of the figure, we give the scale of the maximum
amplitudes corresponding to values of $q$ in the abscissa
(same as in Figure~\ref{fig5}). 
The assumed completeness limit for the OGLE sample of $a=0.3$ 
corresponds to $q \simeq 0.12$.
\label{fig12}}

\clearpage

\begin{deluxetable}{ccccccccccc}
\tabletypesize{\small}
\tablecaption{The amplitude distributions $A^{0.25}_q$ \label{tab1}}
\tablewidth{0pt}
\tablehead{
\colhead{a$\backslash$q}    & \colhead{0.95} & \colhead{0.85} &
\colhead{0.75} & \colhead{0.65} & \colhead{0.55} & 
\colhead{0.45} & \colhead{0.35} & \colhead{0.25} &
\colhead{0.15} & \colhead{0.05}
}
\startdata
0.025 & 0.0916 & 0.0937 & 0.0969 & 0.1006 & 0.1053 & 0.1124 & 0.1227 & 0.1404 & 0.1761 & 0.3309\\
0.075 & 0.0910 & 0.0914 & 0.0930 & 0.0960 & 0.0985 & 0.1030 & 0.1095 & 0.1205 & 0.1376 & 0.2257\\
0.125 & 0.0903 & 0.0906 & 0.0916 & 0.0929 & 0.0942 & 0.0963 & 0.0989 & 0.1029 & 0.1087 & 0.1694\\
0.175 & 0.0838 & 0.0834 & 0.0831 & 0.0829 & 0.0828 & 0.0833 & 0.0836 & 0.0851 & 0.0917 & 0.1327\\
0.225 & 0.0727 & 0.0720 & 0.0718 & 0.0719 & 0.0710 & 0.0709 & 0.0710 & 0.0727 & 0.0938 & 0.1070\\
0.275 & 0.0620 & 0.0613 & 0.0611 & 0.0609 & 0.0605 & 0.0607 & 0.0613 & 0.0645 & 0.1399 & 0.0343\\
0.325 & 0.0541 & 0.0531 & 0.0533 & 0.0535 & 0.0532 & 0.0538 & 0.0548 & 0.0603 & 0.1230 &   0   \\
0.375 & 0.0473 & 0.0477 & 0.0474 & 0.0477 & 0.0478 & 0.0486 & 0.0507 & 0.0652 & 0.0946 &   0   \\
0.425 & 0.0437 & 0.0434 & 0.0432 & 0.0434 & 0.0437 & 0.0451 & 0.0482 & 0.1093 & 0.0345 &   0   \\
0.475 & 0.0402 & 0.0399 & 0.0401 & 0.0402 & 0.0408 & 0.0425 & 0.0482 & 0.1038 &   0    &   0   \\
0.525 & 0.0371 & 0.0369 & 0.0373 & 0.0377 & 0.0384 & 0.0408 & 0.0708 & 0.0687 &   0    &   0   \\
0.575 & 0.0352 & 0.0351 & 0.0353 & 0.0359 & 0.0368 & 0.0410 & 0.0990 & 0.0066 &   0    &   0   \\
0.625 & 0.0338 & 0.0333 & 0.0335 & 0.0344 & 0.0360 & 0.0543 & 0.0703 &   0    &   0    &   0   \\
0.675 & 0.0322 & 0.0317 & 0.0322 & 0.0332 & 0.0369 & 0.0889 & 0.0110 &   0    &   0    &   0   \\
0.725 & 0.0310 & 0.0307 & 0.0311 & 0.0331 & 0.0557 & 0.0558 &   0    &   0    &   0    &   0   \\
0.775 & 0.0304 & 0.0301 & 0.0308 & 0.0359 & 0.0752 & 0.0027 &   0    &   0    &   0    &   0   \\
0.825 & 0.0299 & 0.0297 & 0.0321 & 0.0633 & 0.0231 &   0    &   0    &   0    &   0    &   0   \\
0.875 & 0.0299 & 0.0306 & 0.0508 & 0.0365 &   0    &   0    &   0    &   0    &   0    &   0   \\
0.925 & 0.0317 & 0.0455 & 0.0355 &   0    &   0    &   0    &   0    &   0    &   0    &   0   \\
0.975 & 0.0320 & 0.0198 &   0    &   0    &   0    &   0    &   0    &   0    &   0    &   0   \\
1.025 &   0    &   0    &   0    &   0    &   0    &   0    &   0    &   0    &   0    &   0   \\
\enddata
\tablecomments{The normalized $A^{0.25}_q(a)$ distributions 
(for $f=0.25$) are tabulated in columns for
mass-ratio bins of $\Delta q = 0.1$, centered on the values given
at the top of each column. They have been computed
for the $V$-band, in bins of $\Delta a = 0.05$, 
centered on the amplitude values given in the first column
labeled $a\backslash q$. With the
limited accuracy of the currently available statistics, the distributions
can be used for $V$, $R$ and $I$ photometric bands for stars of F--K 
spectral types.}
\end{deluxetable}


\begin{deluxetable}{cccc}
\tabletypesize{\small}
\tablecaption{The maximum $V$-band amplitude as a function of
$q$ for three values of $f$\label{tab2}}
\tablewidth{0pt}
\tablecolumns{4}
\tablehead{
\colhead{$q$}  & \multicolumn{3}{c}{$f$} \\
      & \colhead{0}    & \colhead{0.25} & \colhead{0.5}    
}
\startdata
  1.00 &  0.945 &  0.989 &  1.031 \\
  0.95 &  0.944 &  0.988 &  1.030 \\
  0.90 &  0.933 &  0.978 &  1.022 \\
  0.85 &  0.918 &  0.962 &  1.005 \\
  0.80 &  0.895 &  0.943 &  0.987 \\
  0.75 &  0.872 &  0.919 &  0.963 \\
  0.70 &  0.845 &  0.892 &  0.938 \\
  0.65 &  0.814 &  0.862 &  0.908 \\
  0.60 &  0.781 &  0.829 &  0.875 \\
  0.55 &  0.745 &  0.793 &  0.839 \\
  0.50 &  0.706 &  0.753 &  0.798 \\
  0.45 &  0.664 &  0.710 &  0.754 \\
  0.40 &  0.618 &  0.663 &  0.706 \\
  0.35 &  0.568 &  0.612 &  0.653 \\
  0.30 &  0.516 &  0.557 &  0.596 \\
  0.25 &  0.458 &  0.496 &  0.532 \\
  0.20 &  0.397 &  0.430 &  0.462 \\
  0.15 &  0.328 &  0.356 &  0.383 \\
  0.10 &  0.252 &  0.272 &  0.292 \\
  0.09 &  0.236 &  0.254 &  0.273 \\
  0.08 &  0.218 &  0.236 &  0.252 \\
  0.07 &  0.200 &  0.216 &  0.231 \\
  0.06 &  0.181 &  0.195 &  0.208 \\
  0.05 &  0.161 &  0.173 &  0.184 \\
  0.04 &  0.140 &  0.149 &  0.158 \\
  0.03 &  0.116 &  0.123 &  0.130 \\
  0.02 &  0.088 &  0.093 &  0.097 \\
  0.01 &  0.054 &  0.057 &  0.059 \\
\enddata
\end{deluxetable}


\begin{deluxetable}{ccc}
\tablecaption{The observed amplitude distributions for the Baade's Window
samples \label{tab3}}
\tablewidth{0pt}
\tablehead{
\colhead{$a$}  & \colhead{BW$_3$}  & \colhead{BW$_5$} 
}
\startdata
 0.025 &  0  &   0 \\
 0.075 &  0  &   0 \\
 0.125 &  8  &  11 \\
 0.175 &  8  &  17 \\
 0.225 & 16  &  44 \\
 0.275 & 16  &  46 \\
 0.325 & 16  &  41 \\
 0.375 &  8  &  20 \\
 0.425 & 13  &  32 \\
 0.475 &  6  &  12 \\
 0.525 &  4  &   8 \\
 0.575 &  0  &   0 \\
 0.625 &  1  &   4 \\
 0.675 &  0  &   0 \\
 0.725 &  0  &   1 \\
 0.775 &  0  &   0 \\
 0.825 &  2  &   2 \\
\enddata
\end{deluxetable}


\begin{deluxetable}{llcc}
\tablecaption{Parameters of the best fitting power law 
distributions $Q(q)$. \label{tab4}}
\tablewidth{0pt}
\tablecolumns{4}
\tablehead{
\colhead{} & \colhead{Parameter}  & \colhead{BW$_3$}  & \colhead{BW$_5$} 
}
\startdata
$Q_a = a_0 (1-q)^{a_1}$ & & & \\
 & $a_0$            &  $214^{+305}_{-135}$  &  $515^{+405}_{-195}$ \\
 & $a_1$            &  $6.0^{+3.8}_{-3.0}$  &  $5.9^{+2.3}_{-1.7}$ \\
 & $\chi^2$         &         5.9           &     11.3            \\
 & $\Sigma A_{pred}$&        301            &     732             \\
 & $\Sigma A_{pred}(a>0.3)$& 44             &     109             \\

\tableline
$Q_b = b_0 q^{b_1}$ & & & \\
 & $b_0$            & \phs $1.8^{+2.5}_{-1.4}$  & \phs $2.7^{+1.8}_{-1.5}$ \\
 & $b_1$            & $-2.0^{+0.7}_{-0.9}$  & $-2.3^{+0.4}_{-0.5}$ \\
 & $\chi^2$         &         6.3           &     12.6            \\
 & $\Sigma A_{pred}$&         850           &     3082            \\
 & $\Sigma A_{pred}(a>0.3)$& 44             &     107             \\

\tableline
Observational data & & & \\
 & $\Sigma A_{obs}$ &        98             &     238             \\
 & $\Sigma A_{obs}(a>0.3)$&  50             &     120             \\
 & \# bins $A(a)$ &           7             &         8           \\
\enddata
\tablecomments{The quality-of-fit measure
$\chi^2$ has been computed for $a>0.3$.
The number of $\Delta a = 0.05$
bins used in the fit is given in the last line of the table;
they can be used to estimate the reduced $\chi^2$ values.
The predicted number of systems over the whole range of the
amplitudes and for $a>0.3$ are given in the lines $\Sigma A_{pred}$
and $\Sigma A_{pred}(a>0.3)$.}
\end{deluxetable}

\clearpage

\begin{deluxetable}{ccccc}
\tablecaption{The expected ratio $N_{pred}/N_{obs}$ 
as a function of $q_{min}$\label{tab5}}
\tablewidth{0pt}
\tablehead{
  & \multicolumn{2}{c}{BW$_3$} & \multicolumn{2}{c}{$a>0.1$} \\  
\colhead{$q_{min}$}  & \colhead{$Q_a$}  & \colhead{$Q_b$} &
\colhead{$Q_a$}  & \colhead{$Q_b$} 
}
\startdata
0.10  & 1.50 & 1.65  & 1.40 & 1.39 \\
0.08  & 1.75 & 2.11  & 1.42 & 1.43 \\
0.06  & 2.03 & 2.87  & 1.46 & 1.49 \\
0.01 &  3.12 & 90.4 & 1.87 & 23.1 \\
\enddata
\tablecomments{The columns labeled BW$_3$ give the ratio of the
total number of contact systems of all amplitudes
to the number observed in the BW$_3$ sample (over the whole
range of amplitudes, not only above 0.3 mag.). 
The columns labeled $a>0.1$ give the expected 
ratio of the total number of systems to the number of systems 
with amplitudes larger than 0.1 mag.}
\end{deluxetable}

\end{document}